\documentclass[preprint,12pt]{aastex}
\usepackage{natbib}
\newcommand{\cmt}{{\rm cm}^{-3}}
\newcommand{\yr}{{\rm yr}}
\newcommand{\pc}{{\rm pc}}

\newcommand{\msun}{M$_\odot$}
\newcommand{\kms}{km s$^{-1}$}
\newcommand{\Mkms}{M$_\odot$ km s$^{-1}$}
\newcommand{\efour}{$\times 10^4$}
\newcommand{\efive}{$\times 10^5$}
\newcommand {\apgt} {\ {\raise-.5ex\hbox{$\buildrel>\over\sim$}}\ }
\newcommand {\aplt} {\ {\raise-.5ex\hbox{$\buildrel<\over\sim$}}\ } 
\shorttitle{Feedback in Disk Galaxies}
\shortauthors{Shetty and Ostriker}

\begin{document}
\title{Cloud and Star Formation in Disk Galaxy Models with Feedback}

\author{Rahul Shetty\altaffilmark{1,2} and Eve C. Ostriker\altaffilmark{1}}
\altaffiltext{1}{Department of Astronomy, University of Maryland, College Park,
  MD 20742} 
\altaffiltext{2}{Harvard-Smithsonian Center for Astrophysics, 60
  Garden Street, Cambridge, MA 02138}
\email{rshetty@cfa.harvard.edu, ostriker@astro.umd.edu}

\begin{abstract}

  We include feedback in global hydrodynamic simulations in order to
  study the star formation properties, and gas structure and dynamics,
  in models of galactic disks.  In previous work, we studied the
  growth of clouds and spiral substructure due to gravitational
  instability.  We extend these models by implementing feedback in
  gravitationally bound clouds: momentum (due to massive stars) is
  injected at a rate proportional to the star formation rate.  This
  mechanical energy disperses cloud gas back into the surrounding ISM,
  truncating star formation in a given cloud, and raising the overall
  level of ambient turbulence.  Propagating star formation can however
  occur as expanding shells collide, enhancing the density and
  triggering new cloud and star formation.  By controlling the
  momentum injection per massive star and the specific star formation
  rate in dense gas, we find that the negative effects of high
  turbulence outweigh the positive ones, and {\it in net} feedback
  reduces the fraction of dense gas and thus the overall star
  formation rate.  The properties of the large clouds that form are
  not, however, very sensitive to feedback, with cutoff masses of a
  few million \msun, similar to observations.  We find a relationship
  between the star formation rate surface density and the gas surface
  density with a power law index $\sim$2 for our models with the
  largest dynamic range, consistent with theoretical expectations for
  our model of disk flaring.  We point out that the value of the
  ``Kennicutt-Schmidt'' index found in numerical simulations (and
  likely in nature) depends on the thickness of the disk, and
  therefore a self-consistent determination must include turbulence
  and resolve the vertical structure.  With our simple feedback
  prescription (a single combined star formation event per cloud), we
  find that global spiral patterns are not sustained; less correlated
  feedback and smaller scale turbulence appear to be necessary for
  spiral patterns to persist.

\end{abstract}

\keywords{galaxies: ISM -- ISM: kinematics and dynamics -- ISM: star
  formation -- turbulence}

\section{Introduction\label{introsec}}
A crucial intermediary for the formation of stars in the ISM is the
gaseous cloud.  Stars form deep within Giant Molecular Clouds (GMCs),
and GMCs themselves may be embedded in larger molecular and atomic
structures, which are referred to as giant molecular associations
(GMAs) and superclouds \citep{Vogel88,ElmElm83}.  The dispersal of
cloud gas, resulting from the ionizing radiation from newly born
stars, stellar winds, and supernovae, limits the lifetimes of GMCs and
therefore determines their net star formation efficiencies.
Supernovae (SN) also play a significant role in maintaining and/or
determining the thermal phase balance of the ISM
\citep{CoxSmith74,McKeeOstriker77,NormanIkeuchi89,deAvillezBreitschwerdt04},
and simple estimates suggest that SN may be the main source of
turbulence, at least in the diffuse ISM
\citep[e.g.][]{Spitzer78,MaclowKlessen04}.  Turbulence in both the
diffuse and dense ISM is in turn considered one of the primary
mechanisms regulating star formation
\citep[e.g.][]{ElmScalo04a,Ballesteros-Paredes07PPV,McKeeOstriker07}.
Since feedback from star formation is linked to the formation,
evolution, and destruction of GMCs, the overall process may be
self-regulating.

The formation and growth of clouds depends on the gravitational
stability of the diffuse gaseous environment.  In disk galaxies,
galactic rotation and thermal pressure, among other factors, act to
oppose the growth of self-gravitating perturbations.  The Toomre $Q$
parameter indicates the susceptibility of axisymmetric perturbations
to grow in uniform thin disks: for $Q<1$, the surface density is
sufficiently large for gas self-gravity to overwhelm the restoring
effects of Coriolis forces and pressure \citep{Toomre64}.  Non-linear
simulations have shown that for non-axisymmetric perturbations, and
including the effects of disk thickness and stellar gravity, the
threshold is $Q\approx1.5$ \citep{KOStone02,KO01,KO07,LiMacKlessen05}.
The observed drop off in star formation activity traced by H$\alpha$ at
large radii supports the idea that stars preferentially form in gravitationally
unstable regions with densities above a critical value
\citep{Kennicutt89,MartinKenn01}.\footnote{Deep H$\alpha$
  \citep{Ferguson98} and UV observations \citep{Thilker07,Boissier07}
  indicate that a fraction of spiral galaxies have extended outer-disk
  star formation, but at much lower levels than in inner disks.  In
  high-$Q$ environments, clouds may grow to become
  self-gravitating by successive inelastic collisions.}  In
general, magnetic fields cannot prevent but only slow the collapse of
gas. In conjunction with other physical mechanisms, magnetic fields
may in fact enhance instability, as is the case when the
magneto-rotational instability \citep[MRI,][]{KOStone03} is present,
or via the magneto-Jeans instability \citep[MJI,][]{KOStone02}.

Star formation must commence soon after the gas accumulates to form
massive clouds, because almost all GMCs contain stars
\citep{BlitzPPV07}.  Ionizing radiation from newly formed stars
subsequently dissociates the molecules, and H II region expansion
disperses the surrounding gas; some fraction of the gas may remain
molecular, but in unbound clouds.  The massive O and B stars reach the
end of their lifetimes in $\sim$2 - 20 Myr, with those over 8 \msun\,
ending as SN. The cumulative effect of feedback from all the
contiguously forming stars contributes to the short estimated GMC
lifetimes of $\sim$20 Myr \citep[e.g.][]{BlitzPPV07}.  The feedback
from star formation could potentially prevent the formation of stars
in nearby regions by driving turbulence and dispersing gas, but it
could potentially trigger collapse events as well.  Collisions between
SN blast waves can result in sufficiently large densities for gas to
collapse and form stars.  It is not yet understood whether (or when)
``positive'' or ``negative'' feedback effects dominate; exploring this
issue is one of the goals of the present work.
  
Despite the host of processes that impact the formation of stars,
observations have shown a clear correlation between the star formation
rate density $\Sigma_{SFR}$ and the gas surface density $\Sigma$, with
power-law forms (in actively star-forming regions)
\begin{equation}
\Sigma_{SFR}\propto\Sigma^{1+p},
\label{KSeq}
\end{equation}
now known as Kennicutt-Schmidt laws \citep{Schmidt59,Kennicutt98}.
Power law indices with $1+p\approx 1-2$ have been found, depending on
whether the total gas mass or just the molecular gas mass is included
in $\Sigma$ \citep[e.g. Bigiel et al. 2008 in
preparation,][]{WongBlitz02,Heyer04,Schuster07,Kennicutt07,Boucheetal07}.
These relations have been identified for a wide range of disk galaxies
at low and high redshifts.  Both global and local versions of the
$\Sigma_{SFR}$ --$\Sigma$ relations have been explored.  In the
former, surface densities are globally averaged within some outer
radius; in the latter, averages are over radial annuli or smaller
regions.  A second empirical law obtained by Kennicutt is
$\Sigma_{SFR}\approx 0.1\Sigma/t_{orb}$, where $t_{orb}$ is the
local orbital time of the gas.

Many theoretical studies have attempted to explain the observed
relations between the star formation rate and the gas surface density.
Simple analytic prescriptions can be obtained that depend on the star
formation efficiency per cloud free-fall time or cloud lifetime, and
yield consistency with the ``orbital time'' empirical relations
\citep{McKeeOstriker07}.  Using global 3D numerical simulations
including gas self-gravity, a prescription for star formation, and
feedback in the form of thermal energy, \citet{TaskerBryan06} found
power law slopes in $\Sigma_{SFR}\propto\Sigma^{1+p}$ similar to
observed values.  \citet{LiMacKlessenLT05,LiMacKlessen06}, using SPH
simulations, found both slopes ($p\sim 0.6$) and normalization similar
to those in \citet{Kennicutt98} ($p\approx 0.4$).  Their simulations
included gravity and sink particles to track the collapsing gas, but
did not treat feedback.  Recently, \citet{RobertsonKravtsov08}
performed simulations that included molecular cooling, and found that
the power law indices obtained by fitting
$\Sigma_{SFR}\propto\Sigma^{1+p}$ are generally steeper if all of
the gas, rather than just molecular gas, is included; this is
consistent with recent observational results.

In this work we investigate how SN driven feedback affects subsequent
star formation in gas disks, including star formation rates.  We model
feedback with a direct momentum input, rather than using a thermal
energy input (when underresolved, the latter approach suffers from
overcooling and the resulting momentum input is too low).  Our work
also differs from other recent simulations in our approach to treating
disk thickness effects; these can be very important to determining the
star formation rate, but direct resolution requiring zones $<5$ pc in
size can be prohibitively expensive to implement in global disk
models.

The evolution of large gas clouds is also relevant to studies of
spiral structure.  In previous work \citep[][hereafter Paper
I]{ShettyOstriker06} we simulated global disks with an external spiral
potential, and found that gravitational instability causes gas in the
spiral arms to collapse to form clouds with masses $\sim 10^7$
M$_\odot$, similar to masses of GMAs and HI superclouds.  We found
that gas self-gravity is also crucial for the growth of spurs (or
feathers), which are interarm features that are connected to the
spiral arm clouds \citep[see also][]{KO02}.  Observations have shown
that spurs are indeed ubiquitous in grand design galaxies, and are
likely connected with large clouds in the spiral arms
\citep{Elm80,LaVigneVogelOstriker06}.  If grand design spiral
structure is long lasting, as hypothesized by density wave theory
\citep[][and references therein]{LinShu64,BertinLin96}, then feedback
mechanisms dispersing the spiral arm clouds must nevertheless leave
the global spiral pattern intact.  One of the goals of this work is to
assess the effect of star formation feedback in massive clouds on the
global spiral morphology.

Conversely, the spiral arms also affect the initial formation of
clouds, therefore also impacting the star formation process.
Observations show that most H$\alpha$ emission in grand design
galaxies occurs downstream from the primary dust lanes.  An
explanation for these observations is that gas is compressed as it
flows through the spiral potential minimum, leading to cloud
formation; then at some later time stars form within these compressed
gas clouds.  Consensus on the exact nature of spiral arm offsets has
not yet been reached, however, owing to both observational limitations
and diverse theoretical views on the star formation process.  Further,
the relative importance of spiral arm triggering is still not
completely understood.  \citet{Vogel88} found that the star formation
efficiency (in molecular gas) in the spiral arms of the grand design
galaxy M51 is only larger by a factor of a few compared with interarm
regions.  Other observational studies comparing star formation rates
in grand design spiral galaxies and those without strong spiral
structure found similar results \citep[see][and references
therein]{Knapen96,Kennicutt98AR}.  As a result, density waves may
primarily gather gas in the spiral arms (enhancing the ability to form
GMCs), but may not significantly affect the star formation efficiency
within any given molecular parcel.  Without a large-scale density
wave, a similar fraction of gas might still collapse (per galactic
orbit) to form clouds via other mechanisms (including large-scale
gravitational instabilities), but not in a coherent fashion.  Here, we
explore the differences in cloud formation properties in gaseous disks
with and without an external spiral driving mechanism.

In this work we are interested in the effect of feedback from star
formation in large clouds, such as GMCs and GMAs, on the star
formation rate, as well as on the overall dynamics and subsequent
cloud formation in galaxy disks.  This work extends the models
presented in Paper I: numerical hydrodynamic simulations of global
disks with gas self-gravity.  With the resolutions of our models,
massive GMAs do not fragment into smaller GMCs, so significant energy
input is required to unbind the gas in these concentrations.  If this
energy is provided by star formation feedback, multiple massive stars
would be needed to destroy the GMAs.  In this work, we model feedback
by considering the impact on large clouds of single energetic events.
In practice, this could represent multiple correlated SN; this can
also be considered simply as an expedient but cleanly parameterizable
feedback model at one extreme of the range of event
correlation\footnote{In future work, we intend to explore how the
  degree of feedback correlation affects the results.}.  We then study
the resulting nature of the turbulent gaseous disk, as well as the
formation and evolution of the clouds that form in the turbulent
medium.  In the next section, we describe our numerical simulation
approach, including model parameters and the feedback algorithm.  We
then present and analyze our simulation results in $\S$\ref{simres}.
In $\S$\ref{discsum} we discuss our results in the context of other
work, and summarize our conclusions.

\section{Modeling Method}

\subsection{Basic Hydrodynamic Equations}

To study the growth and destruction of clouds in a gaseous disk, we
simulate the evolution of the gaseous component by integrating the
equations of hydrodynamics.  As in Paper I, we include the
gravitational potential of the gas.  Our models are two-dimensional,
except that vertical structure of the disk is included in the
calculation of self-gravity, embodied in a function $f(z)$ (see below
and the Appendix).  The governing hydrodynamic equations, including
self-gravity, are:

\begin{equation}
  \frac{\partial \Sigma}{\partial t} + \nabla \cdot (\Sigma {\bf v} ) =
  0,
\label{cont}
\end{equation}
\begin{equation}
  \frac{\partial {\bf v}}{\partial t} + {\bf v}\cdot\nabla {\bf v} +
  \frac{1}{\Sigma}\nabla \Pi =  - \nabla(\Phi_{ext} + \Phi) - \frac{v_c^2}{R}, 
\label{force}\
\end{equation}
\begin{equation}
  \nabla^2\Phi = 4\pi G f(z) \Sigma.
\label{Poisson}
\end{equation}

Here, $\Sigma$, ${\bf v}$, and $\Pi$ are the gas surface density,
vertically averaged velocity, and vertically integrated pressure,
respectively, and $v_c$ is the unperturbed circular orbital velocity.  
For simplicity, we assume an isothermal equation of
state, so that $\Pi = c_s^2\Sigma$, where $c_s$ is the sound speed.
The term $\Phi$ is the gaseous self-gravitational potential.  To grow
gaseous spiral arms, we include an external spiral potential
$\Phi_{ext}$ to model the perturbation produced by the
non-axisymmetric stellar distribution, which is specified at time $t$
in the inertial frame, by

\begin{equation}
\Phi_{ext}(R,\phi;t)=\Phi_{ext,0}\cos[m \phi - \phi_0(R) - m\Omega_pt]
\label{spiralpot}
\end{equation}
where $m$, $\phi_0(R)$, and $\Omega_p$ are the number of arms,
reference phase angle, and spiral pattern speed, respectively.  We
only consider models with a constant pitch angle $i$, so that 
\begin{equation}
\phi_0(r) = -\frac{m}{\tan\,i}\ln(R) + constant.
\label{refang}
\end{equation}

\subsection{Model Parameters}
Similar to Paper I, the sound speed $c_s$ and rotational velocity
$v_c$ are constant in space and time, $c_s = 7$ \kms, and $v_c$ = 210
\kms.  We adopt the code unit of length $L_0 = 1$ kpc.  Using $c_s$ as
the code unit for velocity, the time unit $t_0 = L_0 / c_s = 1.4
\times 10^8$ years, which corresponds to one orbit $t_{orb} = 2
\pi/\Omega_0$ at a fiducial radius $R_0 = L_0 v_c / 2\pi c_s = 4.77$
kpc.  Our results will scale to other values of $R_0$ and $L_0$ with
the same ratio, as well as to models with the same ratio $v_c / c_s =
30$.

In Paper I, we explored different external spiral potential strengths,
\begin{equation}
F\equiv\frac{\Phi_{ext,0}m}{v_c^2\tan i}
\label{eps}
\end{equation}
which is the ratio of the maximum radial perturbation force to the
radial force responsible for a constant rotational velocity $v_c$.  We
found that spurs form in disks with strong external potential
strengths.  Since one of our objectives is to assess the evolution of
the spurs in disks including feedback, here we only simulate disks
with $F$ = 10\%, for both 2 arm and 4 arm spiral galaxies ($m$=2 and
$m$=4).  The corotation radius of 25 $L_0$ corresponds to 25 kpc and a
pattern speed of 8.4 km s$^{-1}$ kpc$^{-1}$, for spiral models using
our fiducial parameters.  We also simulate disks with no external
spiral forcing.

In our computation of gas self-gravity, we include the effect of the
thickness of the disk via $f(z)$, which also acts as softening.  We
assume a Gaussian vertical gas distribution, with scale height $H
\propto R$, so the disk flares at larger radii (see Appendix).  For a
given surface density, the effective midplane density is given by
$\rho_0=\Sigma /( H\sqrt{2\pi})$.  As described in Paper I and
\citet{KO07}, including the effect of thickness provides an important
stabilizing effect on the disk.  For most of our simulations, we use
$H/R$ = 0.01.

As in Paper I, the Toomre parameter $Q_0 \equiv \kappa_0 c_s/(\pi G
\Sigma_0)$ and the surface density $\Sigma_0$ at $R_0$ are related by:
\begin{equation}
\Sigma_0 = \frac{2\sqrt{2} c_s^2 }{G L_0 Q_0 }
=
\frac{32}{Q_0}\, {\rm M}_\odot\, {\rm pc}^{-2}
\left(\frac{c_s}{7\,{{\rm km\, s}^{-1}}}\right)^2
\left( \frac{L_0}{{\rm kpc}}\right)^{-1}.
\label{inits}
\end{equation}
For flat rotation curves, the epicyclic frequency $\kappa = \sqrt2
\Omega = \sqrt2 v_c/R$.  Our models initially have $\Sigma \propto
R^{-1}$, so that $Q$ is constant for the whole disk.

\subsection{Numerical Methods}
Since this work is an extension of previous work, we refer the reader
to Paper I for a description of the cylindrical-symmetry version of
the ZEUS code \citep{StoneNorman92I,StoneNorman92II} that we use to
carry out our simulations.  We use a parallelized version of the
hydrodynamic code and gravitational potential solver, allowing us to
increase the number of zones in the grid relative to the models of
Paper I.  For our standard grid we set the azimuthal range to 0 -
$\pi/2$ radians and the radial range to 4 - 11 kpc.  We implement
outflow and periodic boundary conditions in the radial and azimuthal
directions, respectively.  These models have 1024 radial and 1024
azimuthal zones.  Since the radial grid spacing is logarithmic, the
resolution varies: the linear resolution in each dimension ($\Delta
R$, $R \Delta \phi$) varies from $\sim$(4 pc, 6 pc) in the innermost
region to $\sim$(11 pc, 17 pc) at the outer boundary.  These high
resolutions allow the Truelove criterion \citep{Truelove97} to be
satisfied throughout the simulation as gas collapses to form self
gravitating clumps.

In this work we use a different method to compute the gravitational
potential from that in Paper I.  Here, we use a method derived from
that described by \citet{Kalnajs71} in polar coordinates \citep[see
also][]{BT87}.  This method employs the convolution theorem for a disk
decomposed into logarithmic spiral arcs.  We implement softening to
account for the non-zero thickness of the disk.  We describe the
method in detail in the Appendix.

We note that for simulations with the standard grid and including a
spiral potential, the limit in azimuth requires that $m$=4 (4 arms).
However, we also explore some models with $m$=2 patterns, with the
azimuthal range 0 - $\pi$, using twice as many azimuthal zones than
the standard grid so that the physical resolution of both simulations
are equivalent.

\subsection{Feedback: Event Description and Algorithm \label{feedsec}}
Equations (\ref{cont})-(\ref{Poisson}) only describe the flow as gas
responds to self gravity, and to the external spiral perturbation, if
one is present.  However, those equations do not describe any feedback
that would occur after a self-gravitating cloud forms and fragments
into smaller-scale structures, ultimately forming stars with a range
of masses.  In the real ISM of galaxies, clouds are dispersed by the
combination of photo-evaporation by UV radiation from massive stars,
and the ``mechanical'' destruction by expanding HII regions and SN.

We include in our simulations a very simple feedback prescription by
implementing ``feedback events,'' each representing momentum input
from a number of SN (or, alternatively, multiple overlapping expanding
HII regions).  The specific SN rate, $R_{SN}$, averaged over all mass
$M_{dense}$ above a chosen threshold density in a galaxy is
\begin{equation}
R_{SN} = \frac{{\rm Number\, of\, Supernovae}}{M_{dense} \cdot {\rm time}},
\label{SNrate}
\end{equation}
where $N_{SN}$ is the number of supernovae.  When this rate is applied
to an individual dense cloud of mass $M_{cl}$ with a lifetime
$t_{cl}$, the average number of SN in the cloud will be
\begin{equation}
N_{SN} = R_{SN}\cdot M_{cl}\cdot t_{cl}.
\label{Nsn}
\end{equation}
If the total mass of stars of all masses formed per single SN is
$M_{SN}$, and the star formation efficiency over a cloud lifetime is
$\epsilon_{SF}$, then
\begin{equation}
N_{SN} = \epsilon_{SF} \frac{M_{cl}}{M_{SN}}.
\label{Nsnalt}
\end{equation}
Equating expressions (\ref{Nsn}) and (\ref{Nsnalt}), the mean cloud
lifetime is
\begin{equation}
t_{cl} = \frac{\epsilon_{SF}}{R_{SN}M_{SN}}.
\label{tcloud}
\end{equation}

In a given time interval $\delta t$, such as the time between
successive computations in the numerical evolution, the probability
$P$ that a cloud (of mean lifetime $t_{cl}$) is destroyed is $\delta t /
t_{cl}$.  Thus,
\begin{equation}
P = \delta t \cdot R_{SN} M_{SN} / \epsilon_{SF}
\label{SNprob}
\end{equation}
In our algorithm, clouds are defined as regions above a chosen density
threshold.  If a particular zone is a local density maximum, that zone
is selected as the center of the feedback event.  For any such
identified cloud, a star formation event is initiated with a
probability per timestep given by equation (\ref{SNprob}).  In each
cloud that is determined to undergo feedback, gas is evenly spread out
in a circular region with a prescribed bounding radius.  Gas in each
zone in the circular region is assigned an outward velocity (relative
to the center) to expand the feedback ``bubble.''  A constant
azimuthal velocity is also added such that total galactocentric
angular momentum is conserved.  The velocity profile inside the bubble
is proportional to the distance from the bubble center.  For most of
our simulations, we choose the radius of the feedback bubble to be 75
pc, which corresponds to 12-23 pixels, depending on the radial
location.  In this way, the initially collapsing cloud gas is forced
back into the surrounding ISM.

In our simulation, we only consider the isothermal expansion of the
clouds, since we assume an isothermal equation of state.  Thus, we can
only consider the net energy input at a stage when expansion of the
shell has become strongly radiative.  Numerical simulations show that
for a single SN of energy $E_{SN} \approx 10^{51}$ ergs, the radial momentum
during the radiative stage is $P_{rad} \approx 3-5\times10^{5}$
M$_\odot$ \kms \citep{Chevalier74,Cioffietal88}.  During the
subsequent evolution of the bubble, the shell momentum $P_{sh}$ is
conserved, and is equal to $P_{rad}$. Wind-driven and pressure-driven
H II region bubbles similarly are accelerated to reach a final
momentum $P_{rad}$.

For a total number of massive stars formed given by equation
(\ref{Nsnalt}), and assuming correlation in time, the total momentum
applied to the shell is
\begin{equation}
P_{sh} = N_{SN} P_{rad} = \epsilon_{SF} \frac{M_{cl}}{M_{SN}} P_{rad}
\label{SNmom}
\end{equation}
The shell velocity $V_{sh}$ is
\begin{equation}
V_{sh} = P_{sh}/M_{sh} = \epsilon_{SF} 
\frac{M_{cl}}{M_{sh}} \frac{P_{rad}}{M_{SN}}.
\label{shellv}
\end{equation}
Here, $M_{sh}$ is the sum of $M_{cl}$ and any ambient gas in the
(circular) feedback region.  Assuming $M_{SN}$ = 100 M$_\odot$ and
$\epsilon_{SF}=0.05$, for $P_{rad} = 3\times10^{5}$ \Mkms, $V_{sh}$ =
150 \kms $\times (M_{cl}/M_{sh})$.

Given our feedback prescription, the two key parameters are the
probability per unit time for cloud destruction (eq. [\ref{SNprob}]),
and the momentum input in the feedback event (eq. [\ref{SNmom}]).  For
the simulations presented here, we explore a range in the rate
$R_{SN}$ and in the momentum input per massive star, $P_{rad}$.  The
specific SN rate is set either to $R_{SN} = (10^9\,$ M$_\odot \times
50\,\,{\mathrm {yr}})^{-1}= 2 \times 10^{-11}$ \msun$^{-1} \yr^{-1}$
(comparable to that in the Milky Way), or ten times that rate (these
models are denoted by $R_{SN}=1$ or 10 in Table \ref{standardmods}).
Since $M_{SN}/\epsilon_{SF}$ in equation (\ref{SNprob}) appears as its
inverse in equation (\ref{SNmom}), we fix $M_{SN} = 100 \,$ M$_\odot$
for all simulations, motivated by the initial mass function of
\citet{Kroupa01}, and explore variations in $\epsilon_{SF}$.  
 Scaling to fiducial values, we then have
\begin{equation}
t_{cl} = 2.5 \times 10^7 \yr 
\left(\frac{\epsilon_{SF}}{0.05}\right)
\left(\frac{R_{SN}}{2\times 10^{-11}\, {\rm M}_\odot^{-1}\, \yr^{-1} }\right)^{-1}
\end{equation}
for the typical lifetime of dense clouds.
The momentum $P_{rad}$ is set either to 3.4\efour\ or 3.4\efive\ \Mkms, in
order to allow for a range in feedback energy and out-of-plane losses
(venting from the galaxy) that reduce $P_{rad}$ for a given energy
input. 

We note that for low values of $P_{rad}$, the energy input will not be
sufficient to destroy a dense, bound cloud.  In particular, a cloud of
surface density $\Sigma_{cl}$ will become unbound only if $V_{sh}
\apgt (2G)^{1/2} (\pi \Sigma_{cl} M_{cl})^{1/4}$.  For
$\Sigma_{cl}=200$ \msun $\pc^{-2}$ and $M_{cl}=10^6$ \msun, the
minimum shell velocity is $\sim 15$\kms. For our larger value of
$P_{rad}$, this inequality is comfortably satisfied, but for the
smaller value it is not.  We indeed find that for the low $P_{rad}$
models, clouds are not destroyed by feedback.  For these models, then,
the ratio $t_{cl}\equiv \epsilon_{SF}/(R_{SN}M_{SN})$ becomes the mean
interval between (non-destructive) feedback events in a given cloud..

Before any feedback, the spiral models are executed for some time to
allow gas to concentrate (due to self gravity) and form clouds in
spiral arms.  In simulations without spiral forcing, condensations
begin to grow due to an initial 0.1\% density perturbation.  As a
result of shear, the first structures that form are large scale
flocculent spiral-like features, which we termed ``sheared background
features'' in Paper I.  Gas in these features then collapses to form
distinct clouds.  Thus, we wait until some threshold density is
reached before feedback occurs.  For most models, the threshold
density is $\Sigma/\Sigma_0 = 10$ (this sets the threshold $\Sigma$ at
320 $M_\odot$ pc$^{-2}$ for $Q_0$=1, and 160 $M_\odot$ pc$^{-2}$ for
$Q_0$=2).  We hereafter refer to any contiguous structure in our
simulation with a density above this chosen threshold as a ``cloud,''
regardless of whether the given structure hosts a feedback event or
not.

We note that with our feedback prescription, the star formation rate
is given by the mass in dense gas (i.e. exceeding the threshold
surface density) times $R_{SN} M_{SN}$.  This linear relation is
supported by the shallow slopes of $\Sigma_{SFR}$ versus
$\Sigma_{mol}$ (as observed in CO emission).  In some other recent
work
\citep[e.g.][]{LiMacKlessenLT05,LiMacKlessen06,TaskerBryan06,TaskerBryan08,RobertsonKravtsov08},
the star formation rate is taken as equal to the mass (with density
above some threshold) divided by the free-fall time at that density,
times some efficiency factor.  Our prescription is therefore
equivalent to choosing a ratio of efficiency over free-fall time at
the surface density threshold of $\epsilon_{ff}/t_{ff}= R_{SN} M_{SN}
= (5\times 10^{7-8}\yr)^{-1}$.  Since the mean internal density within
real GMCs (which have surface densities similar to our critical
threshold) is $\sim 100\ \cmt$, with corresponding free-fall time of 4
Myr, our models would cover a range of star formation efficiencies per
free-fall time of $\epsilon_{ff}\sim 1-10$\%.

\section{Simulation Results \label{simres}}

We first present simulations with standard grid parameters, without
spiral structure.  We then show results of simulations including
spiral structure, as well as simulations pertaining to different
radial regions.

Table \ref{standardmods} shows the initial conditions of the standard
set of models we
present, as well as the relevant parameters controlling the feedback
events.  Column (1) lists each model.  Column (2) shows the initial
Toomre $Q$ parameter which is initially constant for the whole disk.
Column (3) indicates the number of arms, all with $F$=10\%.  Column
(4) gives the SN rate, which is required for setting the probability
that a feedback event occurs in a cloud (see eq.  [\ref{SNprob}]).
Column (5) shows the assumed star formation efficiency, and column (6)
gives the adopted momentum input per massive star.  
For these models $H/R=0.01$.

\subsection{Disks without Spiral Structure}

\begin{figure}
\includegraphics[angle=-90,scale=0.75]{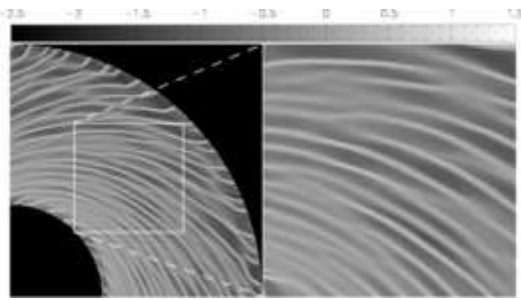}
\caption{Density snapshots of $Q_0=1$ non-spiral model before any
  feedback, at time $t/t_{orb}$ = 0.84.  Gray scale is in units of
  $\log(\Sigma/\Sigma_0)$.}
\label{befSN}
\end{figure}

Figure \ref{befSN} shows a snapshot of model with $Q_0$=1, at time
$t/t_{orb}$ = 0.84, without an external spiral potential and before
any feedback.  As discussed in $\S$\ref{feedsec}, trailing features
grow due to the self-gravity and shear in the disk (see Paper I for
details).  The most dense structures grow as sheared, trailing
features.  It is in these regions where the first SN will occur to
disperse the dense gas.

\begin{figure}[c]
\includegraphics[angle=-90,scale=0.75]{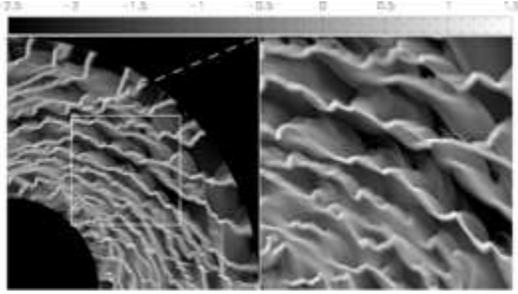}
\caption{Density snapshots of model Q1A, at time $t/t_{orb}$ = 1.125.
  Gray scale is in units of $\log(\Sigma/\Sigma_0)$.}
\label{Q1ASN}
\end{figure}

Figure \ref{Q1ASN} shows a snapshot of model Q1A, at time $t/t_{orb}$
= 1.125.  For model Q1A, the SN parameters are all at the low end of
the range.  At the time of this snapshot, 105 feedback events have
occurred, in clouds which have mean $M_{cl} = 1.2\times10^6$
M$_\odot$.  The main difference between Figures \ref{Q1ASN} and
\ref{befSN} is the shape of the trailing features.  The feedback
events have caused the features to become fragmented at some
locations.  However, dispersal of gas due to feedback was not
sufficient to prevent or reverse the inflow of gas into the high
density agglomerations.  Either the SN do not occur rapidly enough, or
do not have enough momentum to alter the basic morphology.  Even
increasing both the SN rate by a factor of 10, and doubling the star
formation efficiency makes little difference; the strong
self-gravitational force from the trailing features keeps much of the
gas in those structures.  Increasing the SN momentum (or equivalently
the velocity) by up to a factor of 8 still does not significantly
affect the outcome: much of the gas is contained in the sheared
structures at any given time.

\begin{figure}
\includegraphics[angle=-90,scale=0.75]{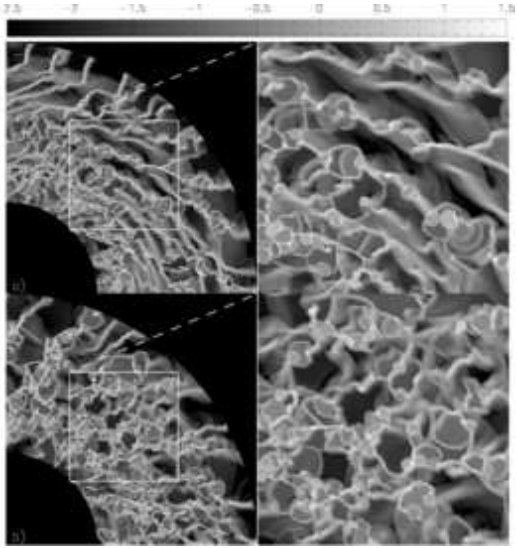}
\caption{Density snapshots of model Q1D, at time $t/t_{orb}$ = 1.125
  (a) and at time $t/t_{orb}$ = 1.375 (b).  Gray scale is in units of
  $\log(\Sigma/\Sigma_0)$.}
\label{Q1BSN}
\end{figure}

It is only when $P_{rad}$ is increased to 3.4\efive\ \Mkms, along with
increasing $R_{SN}$ by a factor of 10 and $\epsilon_{SF}$ to 0.05,
that we find a significant difference compared to the case Q1A, as in
Model Q1D shown in Figure \ref{Q1BSN}.  The velocity is sufficiently
large to drive gas away from the density maxima of the trailing
structures.  Further, the rate is high enough that a large number of
events occur to significantly alter the morphology, in comparison with
Figure \ref{Q1ASN}.  Feedback events in this model are so frequent and
energetic that collisions between bubbles occur.  In some instances,
such collisions create density enhancements that later result in more
collapse and subsequent feedback along the bubble interface.  At time
$t/t_{orb}$ = 1.125 (Fig. \ref{Q1BSN}(a)), we can still make out the
underlying loci of the initial structures formed by gravitational
instability and shear, though 537 feedback events have occurred up to
this time.  Yet, after an additional 26 Myr and 75 feedback events
(Fig. \ref{Q1BSN}(b)), the dominant large scale features do not have a
single pitch angle.  Further, the locations of many of the bubbles are
clustered.  Though gas is driven away from the initial structures
formed before feedback, at later times clouds form in clusters near
the initial density maxima, and where feedback bubbles overlap.
Qualitatively, the features in the disk, consisting of filaments and
bubbles, are similar to the global models including feedback of
\citet{WadaNorman01}.  We discuss the masses of the clouds in both
``non-spiral'' and spiral models in the next section
($\S$\ref{spiralsec}).

In disks with $Q_0=2$, sheared features will also grow, but need more
time to develop than in the $Q_0=1$ disks.  Due to its relative stability,
after $t/t_{orb}=2$ only a few clouds have formed.  As a result,
implementing feedback does not affect the majority of the disk.  To
study the effect of feedback in $Q_0=2$ disks, another mechanism is
necessary to grow clouds everywhere in the disk.  We thus simulate
$Q_0$=2 disks with an external spiral potential, and then implement
feedback to destroy the spiral arm clouds that grow.

\subsection{Disks with Spiral Structure \label{spiralsec}}

In disks with spiral structure, the stellar spiral potential acts as a
source of perturbation; the compression of gas as it flows through the
potential eventually leads to the growth of self gravitating clouds.
We explore the effect of feedback on the morphology of the gaseous
spiral arms and interarm spurs, as well as any subsequent cloud
formation.

\begin{figure}
\includegraphics[angle=-90,scale=0.75]{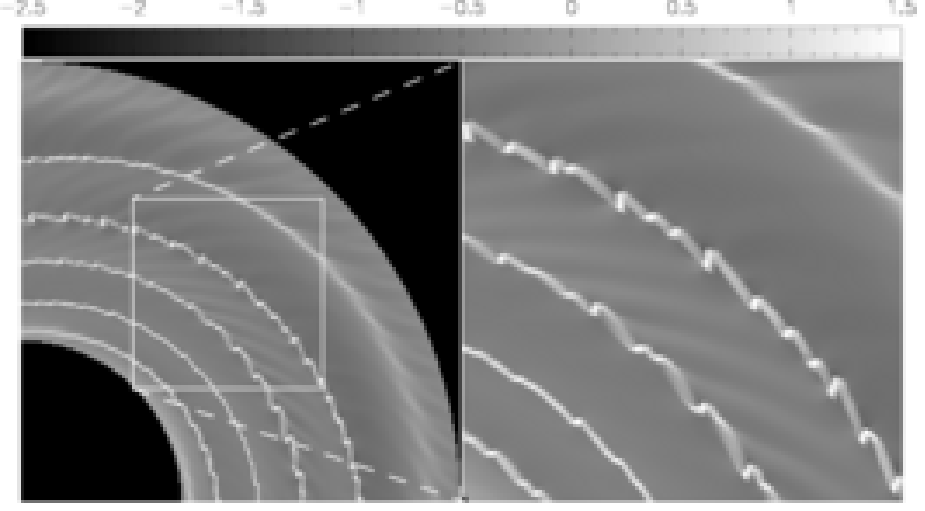}
\caption{$Q_0=1$ spiral model, without feedback, at $t/t_{orb}=0.675$.
  Gray scale is in units of $\log(\Sigma/\Sigma_0)$. }
\label{spmodsQ1}
\end{figure}

\begin{figure}
\includegraphics[angle=-90,scale=0.75]{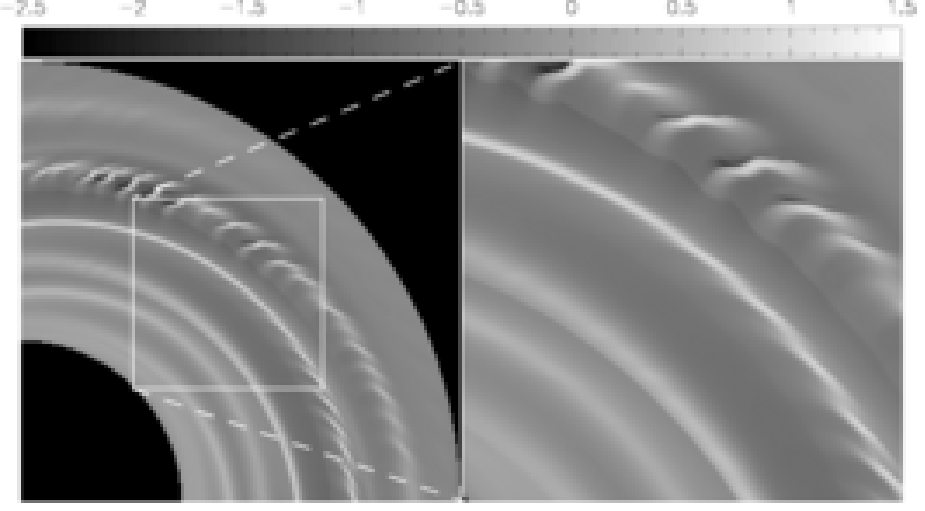}
\caption{$Q_0=2$ spiral model, without feedback, at $t/t_{orb}=1.04$.
  Gray scale is in units of $\log(\Sigma/\Sigma_0)$. }
\label{spmodsQ2}
\end{figure}

Figures \ref{spmodsQ1}-\ref{spmodsQ2} show snapshots of models with
$m=4$, for $Q_0=1$ and $Q_0=2$, without any feedback.  In the spiral
models, the growth of spiral arm clouds occur sooner than clouds
formed by natural instabilities in a rotating self-gravitating
disk.\footnote{In our models, the amplitude of the spiral perturbation
  is ``turned-on'' gradually, reaching the maximum amplitude $F$ at
  $t/t_{orb}$ = 1.  Due to this imposed ``turn-on'' time, the growth
  rate of GMCs in our models is not representative of actual GMC
  formation timescales.}  Figures \ref{spmodsQ1}-\ref{spmodsQ2} show
snapshots of models without any feedback, though in the $Q=1$ snapshot
(Fig. \ref{spmodsQ1}) the densities have surpassed the threshold
density $\Sigma/\Sigma_0=10$ chosen for models with feedback.  Note
that while the $Q=1$ model (with strong self-gravity) shows dense gas
knots within the arm, the $Q=2$ model (with weaker self-gravity) shows
spur-like features; gas does not collapse as promptly.

For both models Q1SA and Q2SA, the SN momenta are insufficient to
offset the growth of clouds and spurs resulting from the spiral
potential.  After a feedback event, the dispersed cloud gas flows back
toward the spiral arm.  As a result, clouds continue to grow over
time.  Further, the spurs also continue to grow in density.  Without
feedback, self-gravitating spiral arm clouds cause the surrounding gas
to flow in with large velocities.  Eventually, the simulations have to
be stopped because the Courant time is too small.  The time when the
simulation ceases, depending partly on our choice of the minimum
acceptable Courant time, also depends on which clouds are (randomly)
selected for feedback; clouds that have produced large inflow
velocities would have to be dispersed for the simulation to continue
to evolve.

\begin{figure}
\includegraphics[angle=-90,scale=0.75]{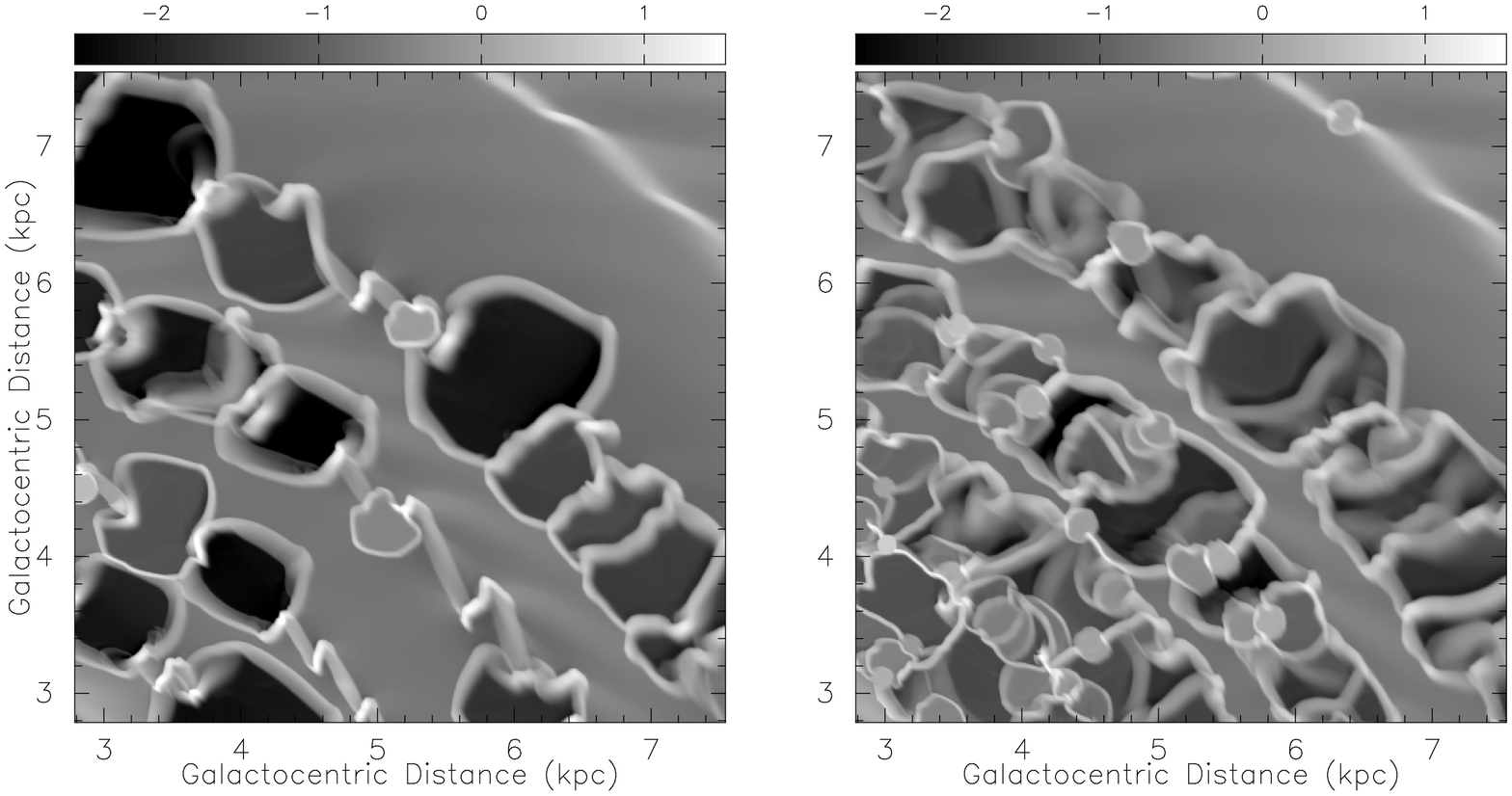}
\caption{Models Q1SC (left) and Q1SD (right) at $t/t_{orb}=0.73$.  
Gray scales are in units of $\log(\Sigma/\Sigma_0)$. }
\label{Q1HV}
\end{figure}

\begin{figure}
\includegraphics[angle=-90,scale=0.75]{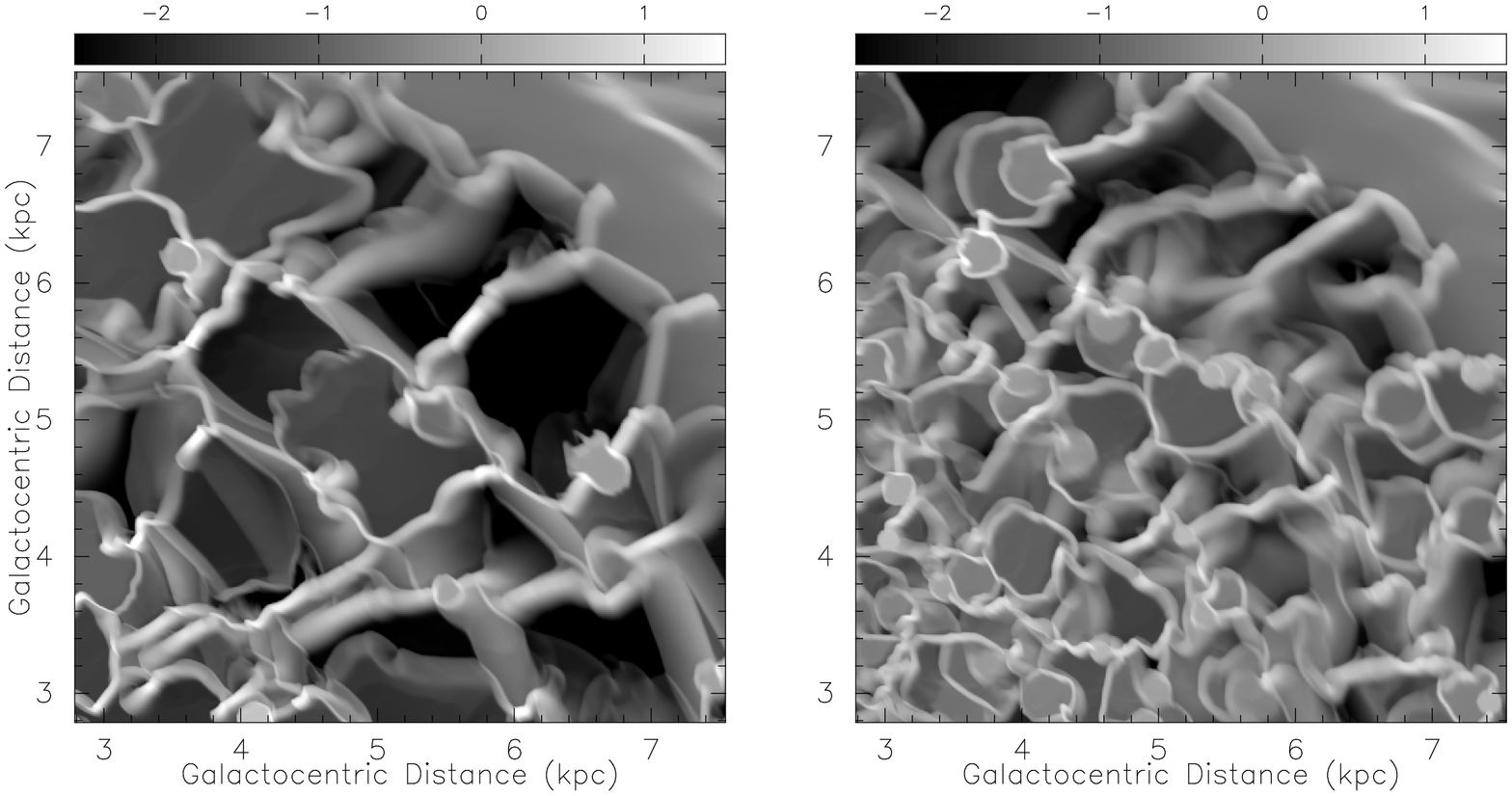}
\caption{Models Q1SC (left) and Q1SD (right) at $t/t_{orb}=1.15$.  
Gray scales are in units of $\log(\Sigma/\Sigma_0)$. }
\label{Q1HVlater}
\end{figure}

We again find that large SN momenta are required to sufficiently
disperse clouds so that immediate re-collapse does not occur.  For
such models, the SN rate has an effect on the number of subsequent
clumps formed.  Figure \ref{Q1HV} shows a snapshot of models Q1SC and
Q1SD, $\sim$ 21 Myr after the first feedback events.  At this time, 53
feedback events have occurred in model Q1SC, and 540 in model Q1SD.
In model Q1SC, it is clear that most, if not all, feedback events
originated in the spiral arms.  However, in model Q1SD, many feedback
events have occurred in interarm regions.  The spiral arms are not as
identifiable, though at this time the remnants of spurs are still
identifiable.  Further, model Q1SD contains many more clumps than
model Q1SC. The enhanced SN rate has caused the collision of more
shell remnants, which lead to formation of self gravitating clumps at
the interfaces.  In both cases, feedback events have caused gas to be
dispersed from the arms, eventually removing any trace of the
underlying spiral potential, as can be seen in Figure \ref{Q1HVlater}.

\begin{figure}
\plottwo{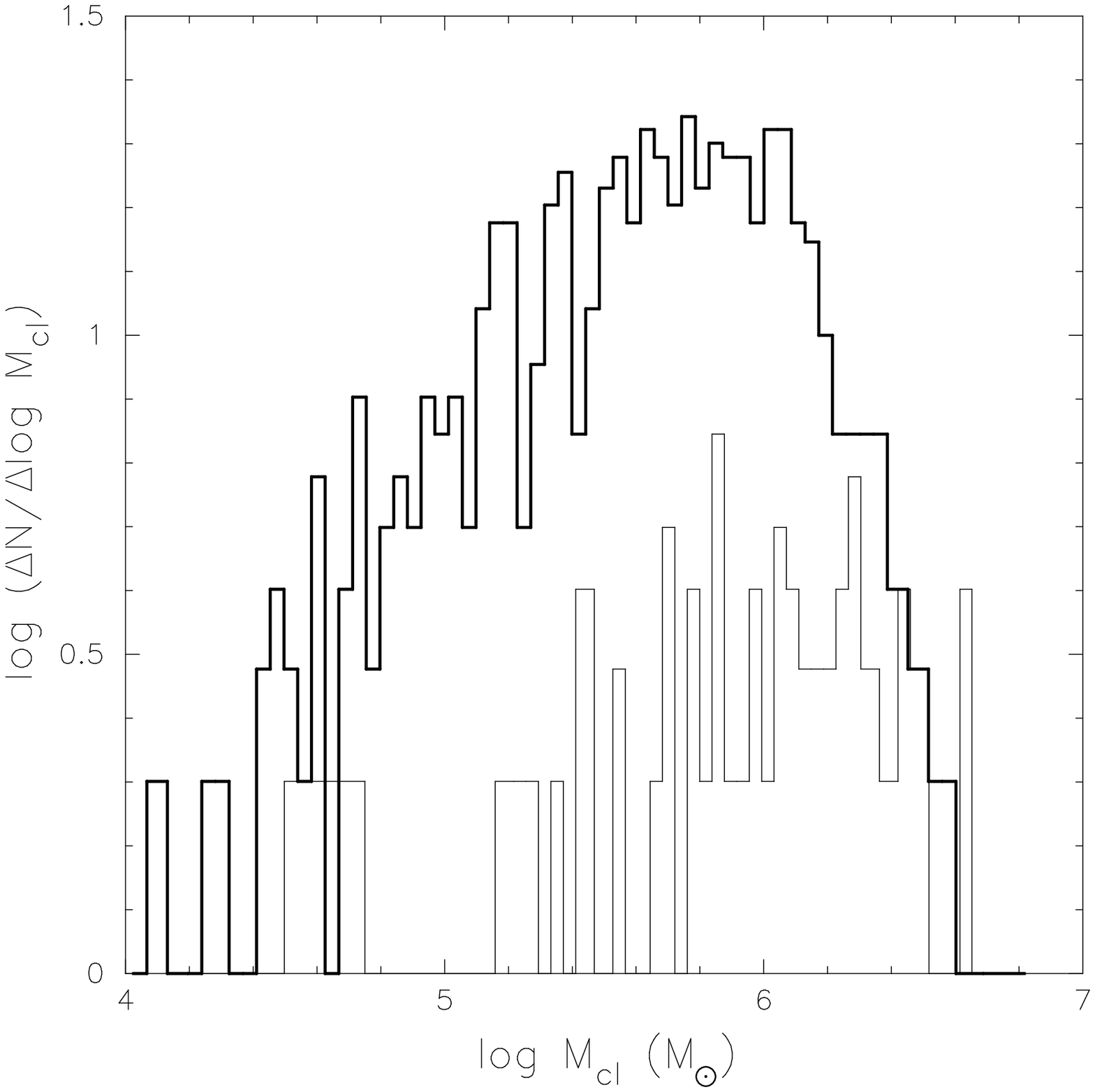}{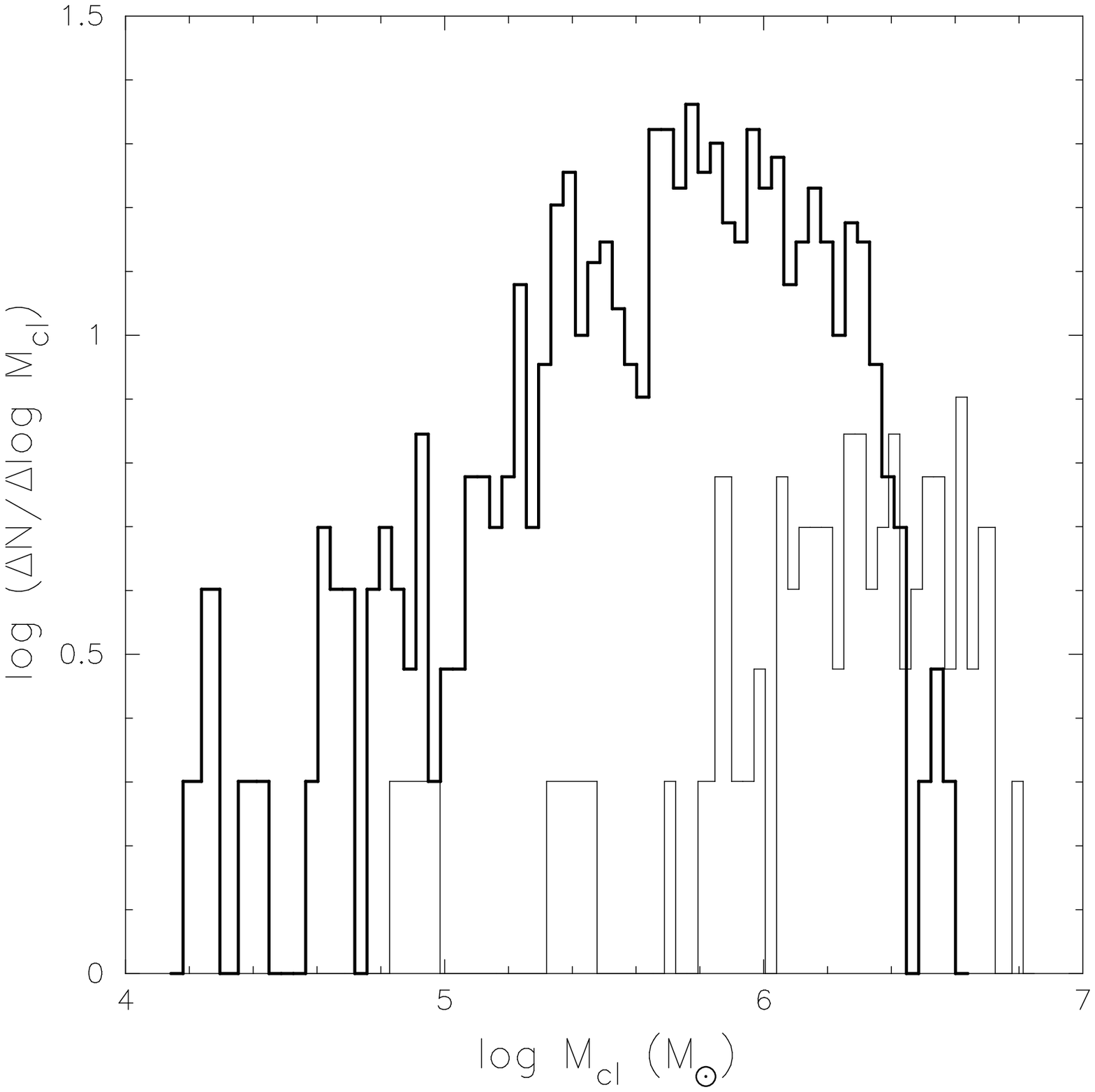}
\caption{Cloud masses in models with strong (thick lines) and weak
  (thin lines) feedback.
Left: Histogram of $M_{cl}$ in models Q1D (thick) and Q1A
  (thin), up until time $t/t_{orb}$ = 1.125 (see Figs. \ref{Q1ASN} -
  \ref{Q1BSN}).  The mean (median) $M_{cl}$ for models Q1A and Q1D are
  1.2$\times10^6$ (0.8$\times10^6$) and 0.7$\times10^6$
  (0.5$\times10^6$) M$_\odot$, respectively. Right: Histogram of
  $M_{cl}$ in models Q1SD (thick) and Q1SA (thin), up until time
  $t/t_{orb}$ = 0.73 (model Q1SD is shown in Fig. \ref{Q1HV}).  The
  mean (median) $M_{cl}$ for models Q1SA and Q1SD are 2.2$\times10^6$
  (1.9$\times10^6$) and 0.8$\times10^6$ (0.6$\times10^6$) M$_\odot$,
  respectively.}
\label{clmasses}
\end{figure}

Figure \ref{clmasses} shows the histogram of the masses of the clouds
$M_{cl}$ that hosted feedback events\footnote{From equation
  (\ref{Nsnalt}), feedback events in model Q1A and Q1D on average
  consist of 300 and 350 SN, respectively.} in models Q1A, Q1D, Q1SA
and Q1SD.  In all cases, the maximum mass of the clouds is below
$10^7$ M$_\odot$, and the means and medians for the distributions lie
in the range $0.5-2.2\times 10^6$ \msun.  In model Q1A, most feedback
events have occurred in the large scale sheared features that grow due
to gravitational instability.  However, in model Q1D, some fraction of
the feedback events have occurred in regions of colliding flows.  The
histogram suggests that clouds formed by colliding flows have
characteristically lower masses than those formed in the large scale
sheared features.  Similarly, in model Q1SA, most feedback events have
occurred in the spiral arms, since most clouds form in the arms.  On
average, the clouds in model Q1SD have lower masses, with many clouds
formed due to colliding flows initiated in earlier feedback events.
For a power law in the mass distribution, $dN/d\log M \propto
M^{-\alpha}$, the distribution in the high end masses for model Q1SD
(below the cutoff at $\log(M)=6.4$) gives $\alpha \sim 0.6$.  This
slope and the upper limit in cloud masses is similar to the range and
the upper limit in the observed masses of GMCs \citep[see][and
references therein]{McKeeOstriker07}.  A histogram of the masses of
all clouds at any given time, normalized by the correct probability
$\delta t/t_{cl}$, reproduces the overall shape of the histogram of
the of masses of clouds with feedback.  A detailed analysis of the
cloud mass distribution is not appropriate here because many of the
lower mass clouds are not well resolved, and because higher-mass
clouds would be subject to turbulent fragmentation that we cannot
follow.  Higher resolution simulations are therefore required to
obtain more complete cloud mass distributions.  Nevertheless, it is
clear that the upper mass limits for clouds in all models are similar
to those in real spiral galaxies.

\subsection{Star Formation Properties}

\subsubsection{Star Formation Rates and Turbulence\label{SFR_turb}}
For comparison to observations, two quantities of interest are the
star formation rate $SFR$ and the turbulent velocity $v_{turb}$. In
each simulation, we record each feedback event to determine the $SFR$.
For some chosen time bin $\Delta t$, we compute
\begin{equation}
SFR = \epsilon_{SF} \frac{\sum M_{cl}}{\Delta t},
\label{SFR}
\end{equation}
where $\sum M_{cl}$ is the total mass of all gas in clumps (i.e. above
the chosen threshold surface density) that have undergone feedback
events in the chosen time interval. (Recall that the mean lifetime of
clouds, or the mean interval between star formation events if they are
non-destructive, is given by eq. [\ref{tcloud}].)

\begin{figure}
\includegraphics[angle=-90,scale=0.5]{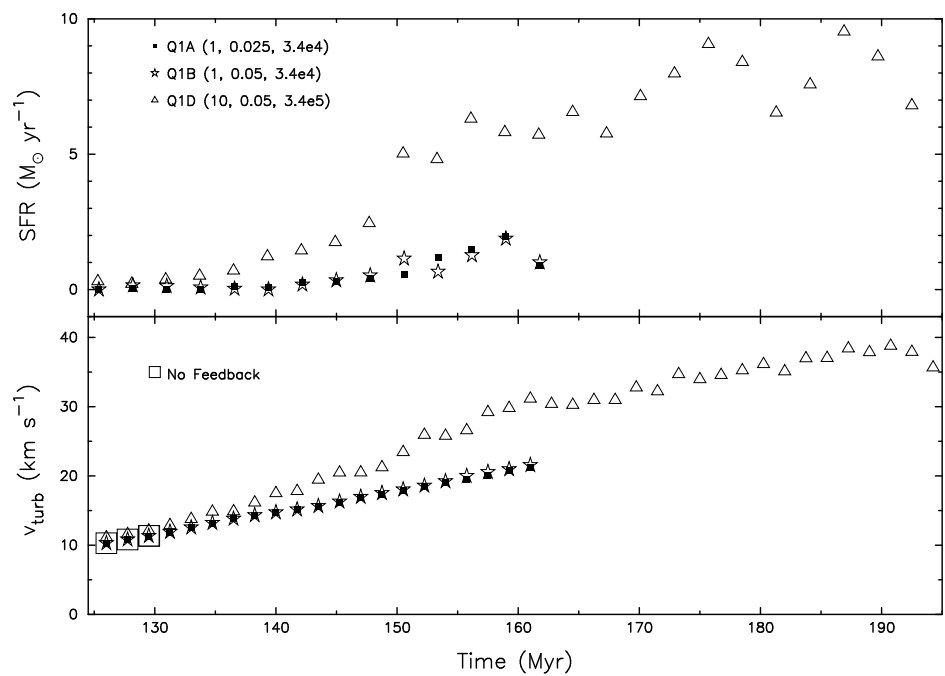}
\caption{$SFR$ (top) and $v_{turb}$ (bottom) for models with $Q_0$=1,
  without spiral structure.  The values in parentheses in the legend
  are the SN rate parameter $R_{SN}$, the star formation efficiency
  $\epsilon_{SF}$, and SN momentum $P_{rad}$ (in \Mkms) of each model.
  The large open squares in the bottom panel are the turbulent velocities
  for a simulation without any feedback.}
\label{Q1SFR}
\end{figure}

We define the turbulent velocity as the RMS sum of any non-circular
velocities, weighted by the corresponding mass:
\begin{equation}
v_{turb} = \left ( \frac{\sum (\delta{\bf
    v}_{i,j})^2\Sigma_{i,j} A_{i,j}}{\sum\Sigma_{i,j} A_{i,j}} \right)^\frac{1}{2},
\label{vturb}
\end{equation}
where $A_{i,j}$ is the area of each zone and only non-circular
velocity components are considered: $\delta{\bf v} = {\bf v} - v_c
{\bf \hat\phi}$.  Figure \ref{Q1SFR} shows the star formation rate and
turbulent velocity as a function of time, for the $Q_0$=1 models
without spiral structure.  The time bin $\Delta t$ for our $SFR$
calculation is 3 Myr.  In these models, the first feedback events
occur at time $\sim$125 Myr.  However, for the first $\sim$25 Myr
after feedback begins, the $SFR$ for all models is only a few
M$_\odot$ yr$^{-1}$.  Only $\sim$25 Myr after the first feedback
events does the $SFR$ substantially increase, owing to ``propagating''
star formation.  Further, the Q1D model with large feedback momenta
($P_{rad}$ = 3.4\efive\ \Mkms) and large SN rate ($R_{SN}=10$) has the
$SFR$ increase to $\sim$10 M$_\odot$ yr$^{-1}$.  This occurs because
with large velocities and a high global rate, adjacent shells collide
and more clouds are formed in the interfaces, which may subsequently
undergo star formation.

The bottom panel of Figure \ref{Q1SFR} shows $v_{turb}$, for all
feedback models without spiral structure, together with results from a
simulation without any feedback.  For the later case, we just allow
self-gravity to grow clouds indefinitely.  When we compute $v_{turb}$
in the model without feedback considering only the low density gas, we
obtain similar values.  This suggests that, before any feedback,
large-scale motions from disk self-gravity and shear are the primary
sources of turbulence \citep[see][]{KO07}.  The models with low
feedback momentum continue the trend of $v_{turb}$ established by the
no-feedback case.  In a few instances of enhanced feedback, there is a
corresponding jump in $v_{turb}$.  The enhanced $SFR$ at later times
for the model with large SN momenta also increases levels of
$v_{turb}$.

\begin{figure}
\includegraphics[angle=-90,scale=0.5]{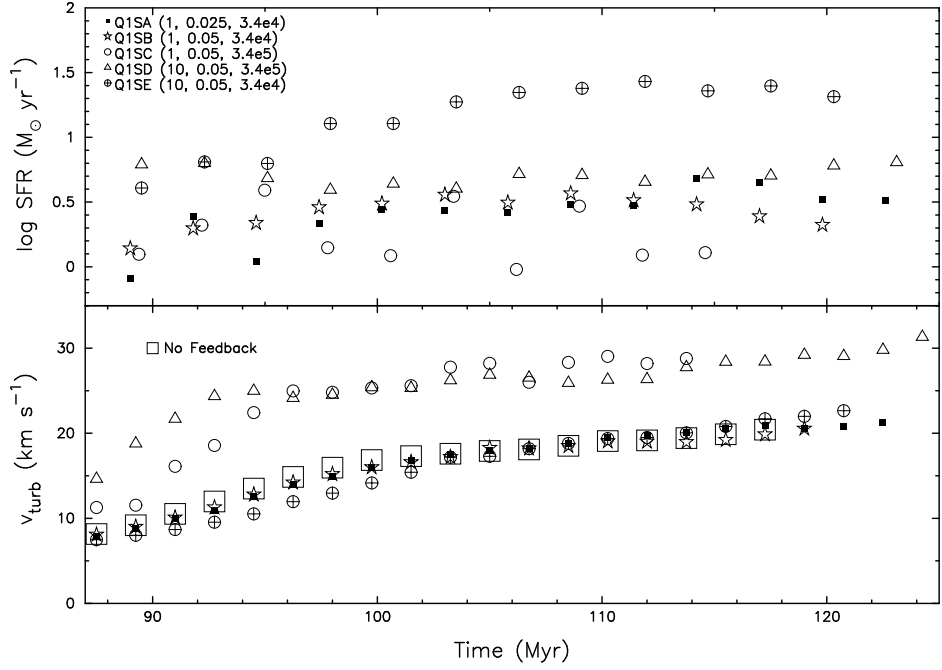}
\caption{$SFR$ (top) and $v_{turb}$ (bottom) for models with $Q_0$=1,
  as in Figure \ref{Q1SFR}, but with spiral structure.}
\label{Q1SSFR}
\end{figure}

\begin{figure}
\includegraphics[angle=-90,scale=0.5]{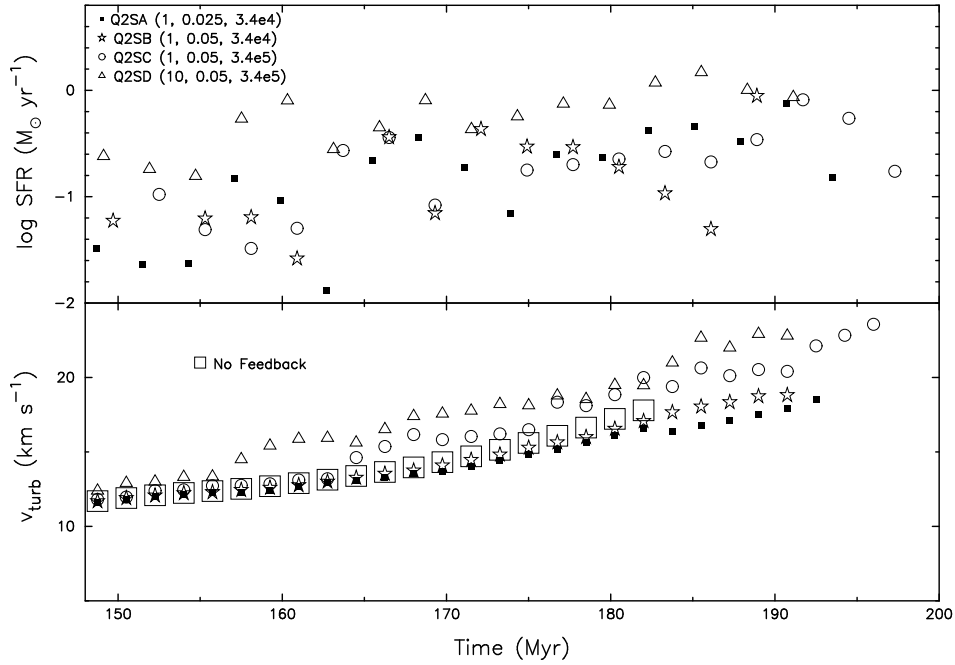}
\caption{$SFR$ (top) and $v_{turb}$ (bottom), as in Figure
  \ref{Q1SSFR}, but for models with $Q_0$=2, with spiral structure.}
\label{Q2SSFR}
\end{figure}

Figures \ref{Q1SSFR} and \ref{Q2SSFR} show the $SFR$ and $v_{turb}$
for the spiral models with $Q_0=1$ and $Q_0=2$.  Comparing Figures
\ref{Q1SSFR} (with spiral structure) and \ref{Q1SFR} (without spiral
structure), the star formation rate is consistent to within a factor
of 2, although slightly larger in some of the spiral models.  The
general trends from the models without spiral structure are reproduced
in Figures \ref{Q1SSFR} and \ref{Q2SSFR}.  Earlier times are shown in
Figure \ref{Q1SSFR}, since the spiral arms cause gas to collapse into
clouds sooner.  It is clear that only in models with large SN shell
velocities -- and regardless of the input rate $R_{SN}$ --
do the turbulent velocities increase appreciably;
otherwise, the turbulent velocity (as we have defined it) is dominated
by effects from gas self-gravity.

It is interesting to compare results from pairs of models in which one
parameter is varied and the others are controlled.  Comparing models
Q1SE and Q1SB, both have the same $\epsilon_{SF}=0.05$ and
$P_{rad}=3.4 \times 10^4$ \Mkms, but the former has $R_{SN}$
larger by a factor 10.  The measured SFR in Q1SE is a factor $\sim 10$
larger than that in Q1SB, consistent with the naive expectation that
$SFR \propto R_{SN}$.  However, when we compare Q1SD with Q1SC, which
again differ in $R_{SN}$ by a factor 10, we find SFR ratios differing
only by a factor $\sim 4$.  This same trend is also true for models
Q2SD and Q2SC.  The reason for this difference in dependence on
$R_{SN}$ is that the E and B models have low $P_{rad}$ and low
turbulence levels, whereas the C and D models have higher $P_{rad}$
and turbulence.  Thus, stronger feedback causes the scaling of SFR to
depart from $SFR \propto R_{SN}$.  We note that since $SFR = R_{SN}
M_{SN} M_{dense}$ by definition, the ratios of specific SFR between any two
models differ by their ratios of $R_{SN} M_{dense}/M_{tot}$.  Thus,
if SFR increases at a rate less than $\propto R_{SN}$, it implies that
increasing $R_{SN}$ {\it decreases} the dense gas fraction $M_{dense}/M_{tot}$.

We can directly investigate the effect of turbulence by comparing the
pair Q1SB and Q1SC, which have the same $\epsilon_{SF}=0.05$ and
$R_{SN}$, but momentum input parameters differing by a factor 10.  As
noted above, this increases the turbulence level in Q1SC compared to
Q1SB by about 10 \kms.  It also reduces the SFR in Q1SC compared to
Q1SB, by a factor $\sim 2-4$.  Similarly, Q1SE has lower $P_{rad}$
than Q1SD, and a substantially lower turbulence level.  For this pair,
too, the SFR in the lower-turbulence model is higher by a factor $\sim
3-5$.  As discussed in \S \ref{introsec}, in principle turbulence
could both enhance star formation (by creating more dense gas in
compressions), and suppress star formation (by destroying overdense
structures with rarefactions and shear flows).  Examining the
evolution of Q1SE indeed shows that feedback events only slightly
expand clouds, and collapse subsequently resumes.  On the other hand,
clouds in model Q1SD are completely destroyed after a single feedback
event.  Evidently, in the models with strong feedback-driven
turbulence, the rate of new cloud formation from shell collisions does
not compensate for the truncation of star formation when a given cloud
is destroyed.

The comparisons of (Q1SB,Q1SC) and (Q1SE,Q1SD) indicate that {\it in
  net}, the increase of turbulence reduces star formation.\footnote{We
  note that models Q2SB and Q2SC also show the same generic behavior,
  but with a smaller difference in the turbulence level, the
  suppression of star formation is also lesser.}  Since the specific SFR is
proportional to the dense gas fraction if $R_{SN}$ is held fixed,
these results imply that the dense gas mass fraction is lower when the
turbulence level is higher.

\begin{figure}
\includegraphics[angle=-90,scale=0.5]{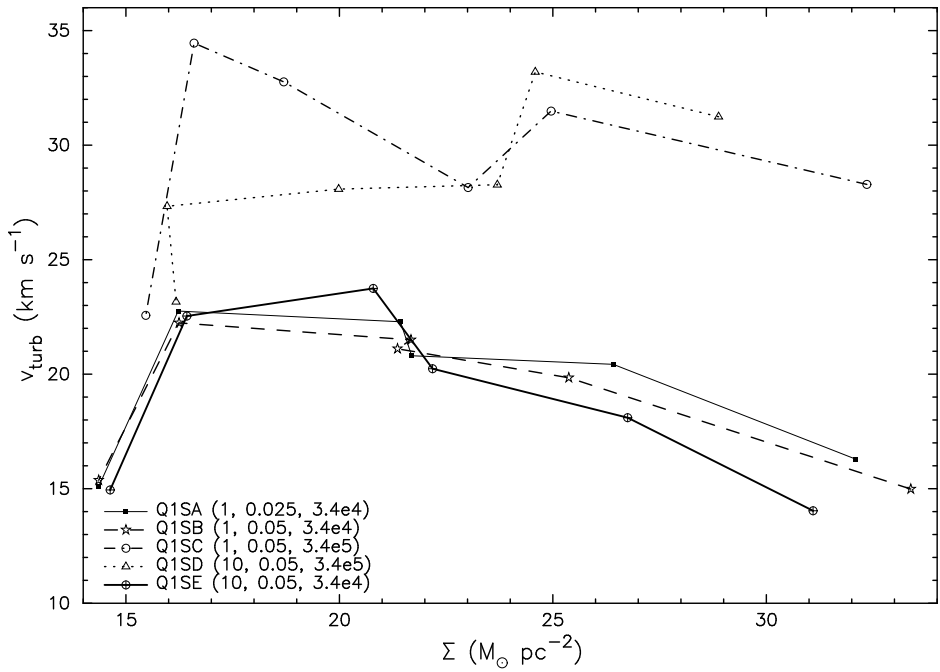}
\caption{Mass weighted turbulent velocities vs. mean surface density
  of Q1S models, averaged in annuli of widths 1 kpc, and in the time
  interval $t/t_{orb} \in 100 - 116$ Myr.  Turbulent velocities are
  only shown from annuli and time intervals within which feedback
  events have occurred.}
\label{dispsig}
\end{figure}

We show the relationship between the mass weighted turbulent velocity
and the surface density in Figure \ref{dispsig}.  Most feedback events
occur in high density regions.  In the higher density regions, the
difference in turbulent velocities (or the velocity dispersions)
between models with $P_{rad}$ = 3.4\efour\ \Mkms\ and $P_{rad}$ =
3.4\efive\ \Mkms\ is \apgt 7 \kms.  At lower density regions, where
there have been fewer feedback events, the dispersions of all models
are comparable.

\begin{figure}
\includegraphics[angle=-90,scale=0.5]{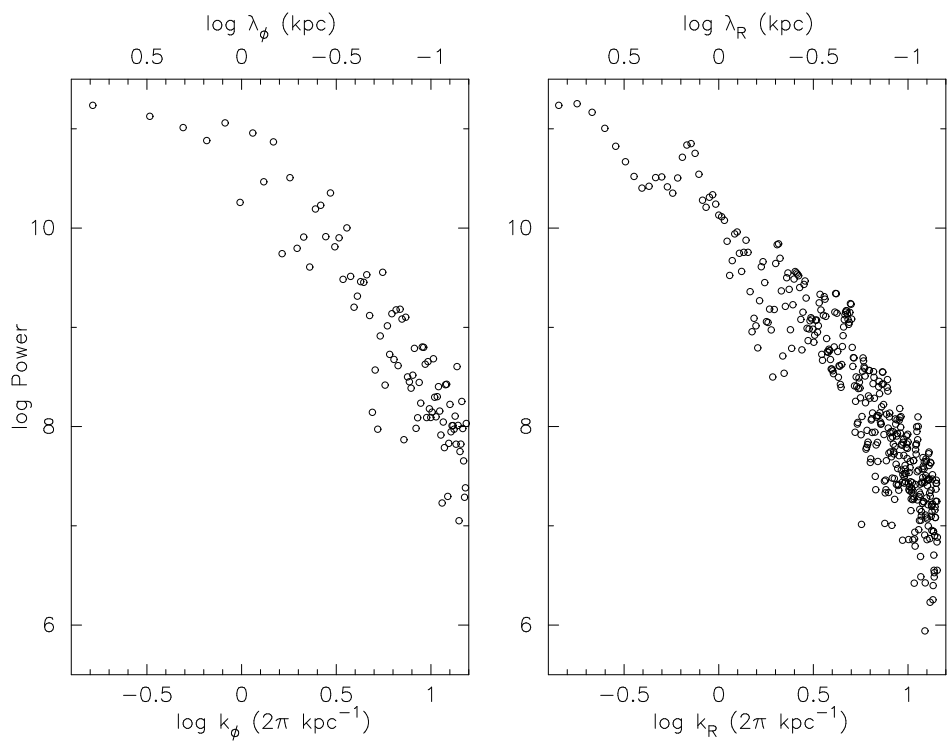}
\caption{Turbulent power spectra of Model Q1D. Power is shown at
  constant $k_R$ (left) and constant $k_\phi$ (right) (each slice is 
  along the minimum nonzero value of the respective $k$).  To obtain the
  dimensions of $k_\phi$ ($k_\phi = m n_\phi/R$), we use the mean
  radius of the grid.  Best fits for values between $\log(k) \in 0 -
  1$ gives slopes of -2.9 (left) and -2.4 (right).}
\label{turbspec}
\end{figure}

Figure \ref{turbspec} shows the turbulent power spectrum (power
$\propto v^2$) of model Q1D.  The power is shown at constant
wavenumbers $k_R$ and $k_\phi$.  The slopes of the power spectra range
from -2.5 to -3.  For models that evolve for significant amounts of
time, such as model Q1D, the power spectra are relatively independent
of time.  These results are consistent with turbulence dominated by
numerous shocks, or Burgers turbulence.  From Figure \ref{turbspec},
the amplitudes of turbulence evidently decrease at smaller scales.

The total turbulent amplitudes shown in Figs.
\ref{Q1SFR}-\ref{Q2SSFR} represent the velocity dispersion averaged
over the whole disk.  For the purposes of assessing turbulent
contributions to local disk stability, however, only the level of
turbulence within a Jeans length $\sim c_s^2/(G \Sigma)$ is relevant.
Furthermore, local observations of turbulence within the Milky Way
generally measure velocity dispersions on scales less than the disk
thickness.  Thus, it is useful to estimate the turbulent amplitudes at
smaller scales than the whole disk.  We do this by running a window
(or ``beam'') of 1 kpc or 100 pc over the map, and finding the
dispersion of the velocity within this window at locations separated
by the window size. When all zones within the window are weighted
equally (as is true for the velocity power spectrum), we find that the
mean velocity dispersions for model Q1D on scales of 1 kpc and 100 pc
are 18 \kms\ and 6.5 \kms, respectively.  When we weight by mass, the
respective velocity dispersions are 31 \kms\ and 10 \kms.  The larger
values obtained when weighting by mass are indicative of the
importance of dense expanding shells in driving the turbulence.

Since turbulence adds to the total momentum flux (the ram pressure
acts similarly to the thermal pressure), a common
assumption is that the sound speed $c_s$ can be replaced by
\begin{equation}
c_{eff}^2 = c_s^2 + \sigma_R^2
\label{c_effeqn}
\end{equation}
in the dispersion relations that characterize stability to
axisymmetric modes, where $\sigma_R$ is the radial component of the
velocity dispersion.  For models Q1SC and Q1SD, which have high
$P_{rad}$, we find that the mean values of $\sigma_R$ on kpc scales
are 17 and 18 \kms, respectively.  For the corresponding models Q1SB
and Q1SE that have low $P_{rad}$, on the other hand, the values of
$\sigma_R$ on kpc scales are 10 and 8 \kms, respectively.  Thus, the
values of $c_{eff}$ exceed $c_s$ by a factor 1.6 for the
low-turbulence models, whereas this increases to a factor 2.7 for the
high-turbulence models.  Our results discussed above indicate a
decrease in the star formation rate with increasing $c_{eff}$; we
discuss theoretical ideas related to this finding in \S\ref{SFpredict}
below.

\subsubsection{Kennicutt-Schmidt Law}
Figure \ref{QKS} shows the local star formation rate per area as a
function of mean surface density.  To obtain these points, simulation
data were binned in radius and time, of widths 1 kpc and 18 Myr,
respectively.  Only models with a sufficient number of points, which
is dependent to some degree on the number of feedback events, are
shown.  Best fit lines to the data points are also shown.  The rates
show considerable scatter, both between models with different
parameters, as well as among points from a given model.  However,
where a large dynamic range is available, as is the case for the Q1D
model and its extension to smaller radii (see below), a power law
relation $\Sigma_{SFR} \propto \Sigma^{1+p}$ is quite clear.

The $Q_0=1$ models, both with and without a spiral perturbation, and
with different feedback parameters, generally give slopes
$1+p\sim1-3$.  Most of the $Q_0=1$ models evolve for sufficiently long
times that gas in the first clouds that are formed are allowed to be
recycled into subsequently formed clouds several times.  The $Q_0=2$
models, on the other hand, give a variety of slopes, and the
relationship between the star formation rate and surface density is
not as well correlated as in the $Q_0=1$ models.  For the $Q_0=2$
models, the number of feedback events is insufficient to affect much
of the disk.  As a result, some clouds continue to collapse, and the
Courant condition would demand an extremely small time step; at this
point, we halt the simulation.  Since the stochastic feedback events
do not result in developed turbulence and a steady state is not
approached in the $Q_0=2$ models, the SFR as computed is sensitive to
model parameters governing the feedback events.

For some of our models, we have also run simulations of the inner
regions of disks, with radial extent $R\in0.8-2.2$ kpc.  The other
parameters are the same as for the standard models.  The only
difference here, besides the radial range, is the initial surface
density.  Since $Q_0$ is constant, and $\Sigma_0\propto (Q_0 R)^{-1}$,
the initial surface density at all radii is increased by a factor 5
compared to the standard models with $R\in4-11$ kpc.

\begin{figure}
\plottwo{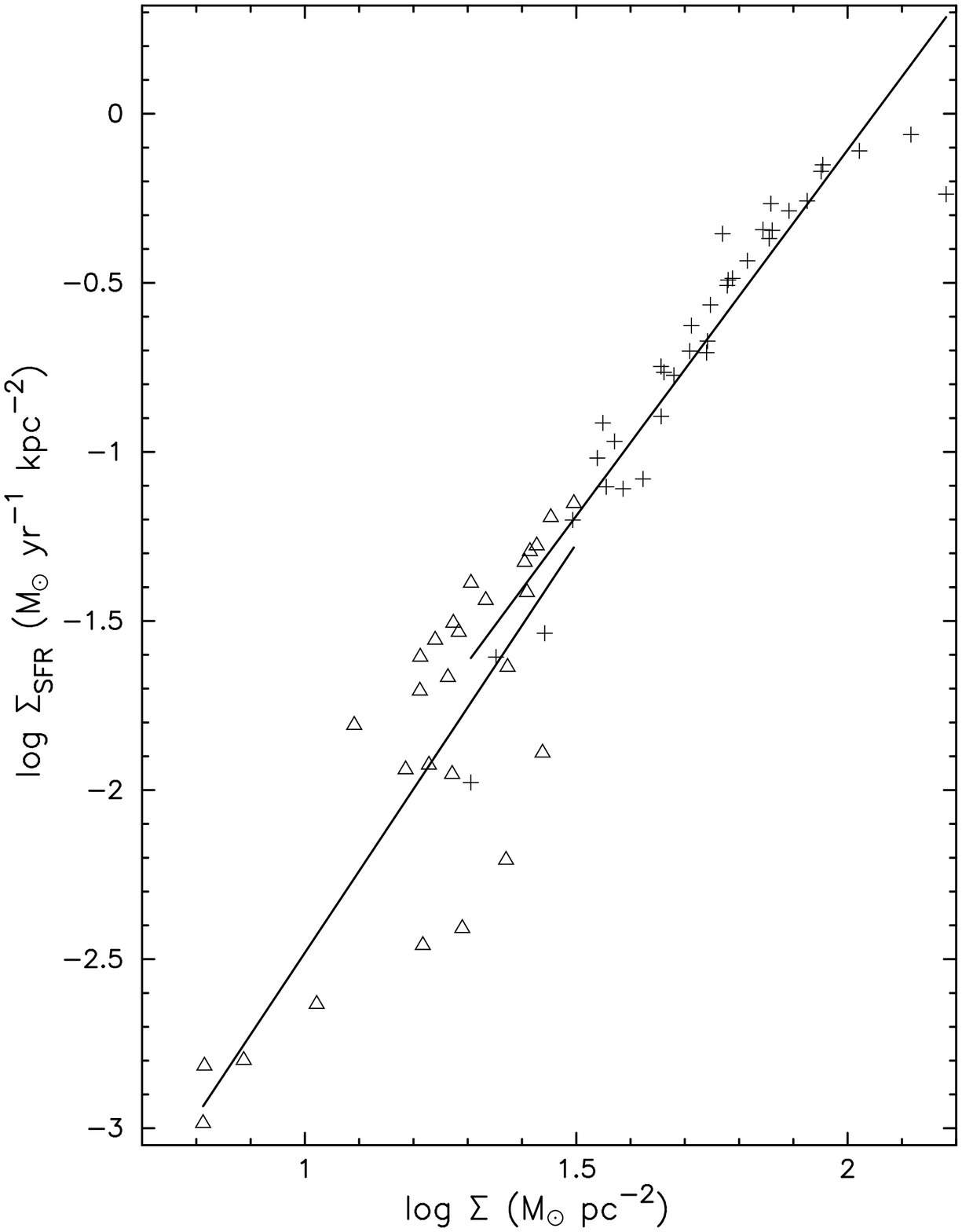}{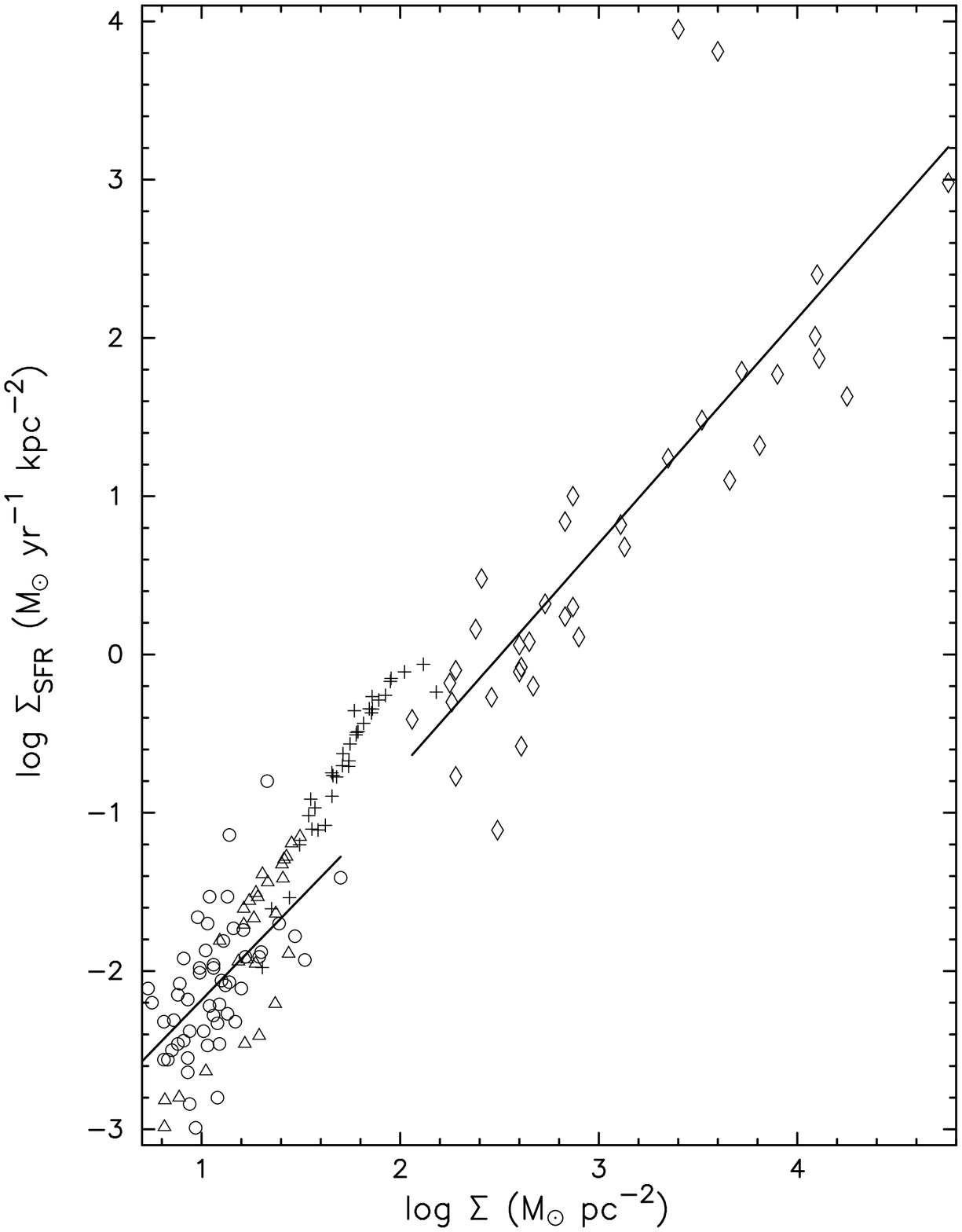}
\caption{Left: Star formation rate vs. surface density for model Q1D
  (triangles; $R\in$ 4 - 11 kpc), as well as the corresponding model of
  the inner region (crosses; $R\in$ 0.8 - 2.2 kpc).  Best fit lines
  for each model are also shown, with slopes of 2.4 for the 4 - 11 kpc
  model and 2.2 for the 0.8 - 2.2 kpc model.  Right: Triangles and
  crosses from figure on the left are shown, along with globally
  averaged observational data from \citet{Kennicutt98}: circles show
  normal spirals, with best fit slope of 1.3, and diamonds show IR starburst
  sources, with best fit slope of 1.4.}
\label{innercomp}
\end{figure}

Figure \ref{innercomp} shows the star formation rate as a function of
surface density for model Q1D together with the corresponding inner
region model.  The larger surface density does indeed lead to higher
star formation rates, with a slope $1+p=2.2$ that is similar to the
value $1+p=2.4$ of the standard model.  We find similar trends for
other inner disk models in comparison with the corresponding standard
models.  For comparison, Figure \ref{innercomp} also shows data from
\citet{Kennicutt98}.  Each point indicates the globally averaged star
formation rate for individual galaxies or their central regions (for
starbursts).  Though there is less scatter in the simulation points,
the slope of the $\Sigma_{SFR}$ - $\Sigma$ relation from the
simulations ($\sim$2.3) is larger than the slope from observational
data ($1+p\sim$1.4).  At the low $\Sigma$ end, the model results
overlap with the observed points.

\subsubsection{Predicting  Star Formation Times\label{SFpredict}}
The star formation (or gas depletion) time $t_{SF}$ for the whole
gaseous component of a galaxy is the time required for all the gas to
be converted to stars if the star formation proceeds as it has been
during a given interval $\Delta t$:
\begin{equation}
t_{SF} = \Delta t \frac{M_{tot}}{\epsilon_{SF} \sum M_{cl}},
\label{deptime}
\end{equation}
where $M_{tot}$ is the total mass in a given annulus.  This quantity
can be measured in our simulations; the summation in equation
(\ref{deptime}) is taken over all clouds in which a feedback event has
occurred, as in equation (\ref{SFR}).

Observationally, if the SN rate per dense gas mass $R_{SN}$ is known,
the star formation (or gas depletion) time can also be estimated based on
the total amount of gas and the portion in dense clouds as:
\begin{equation}
t^\prime_{SF} = \frac{M_{tot}}{M_{dense} M_{SN} R_{SN}},
\label{deptimeapp}
\end{equation}
where $M_{dense}$ is the total mass of gas above some chosen threshold
density.  Since $R_{SN}M_{SN} = \epsilon_{SF} / t_{cl}$ from equation
(\ref{tcloud}), the results of equations (\ref{deptime}) and
(\ref{deptimeapp}) should agree on average.  With our two parameter
choices $R_{SN}$ = 1 or 10 (in units 2 $\times 10^{-11}$ \msun$^{-1}$
$\yr^{-1}$), this implies $t^\prime_{SF} = (1\ {\rm or}\ 0.1) \times
(M_{tot}/M_{dense}) \times 5\times10^8\ \yr$.  We note that if the
star formation or gas depletion time were computed only for
\textit{dense} gas (with local surface density $\apgt 200$ \msun\
$\pc^{-2}$), then for our prescription it would simply be equal to a
constant, $t^\prime_{SF}(dense)=(M_{SN} R_{SN})^{-1}=5\times10^{7}$ or
$5\times10^{8}\ \yr$ for $R_{SN}$ = 10 or 1, respectively.

\begin{figure}
\includegraphics[angle=-90,scale=0.5]{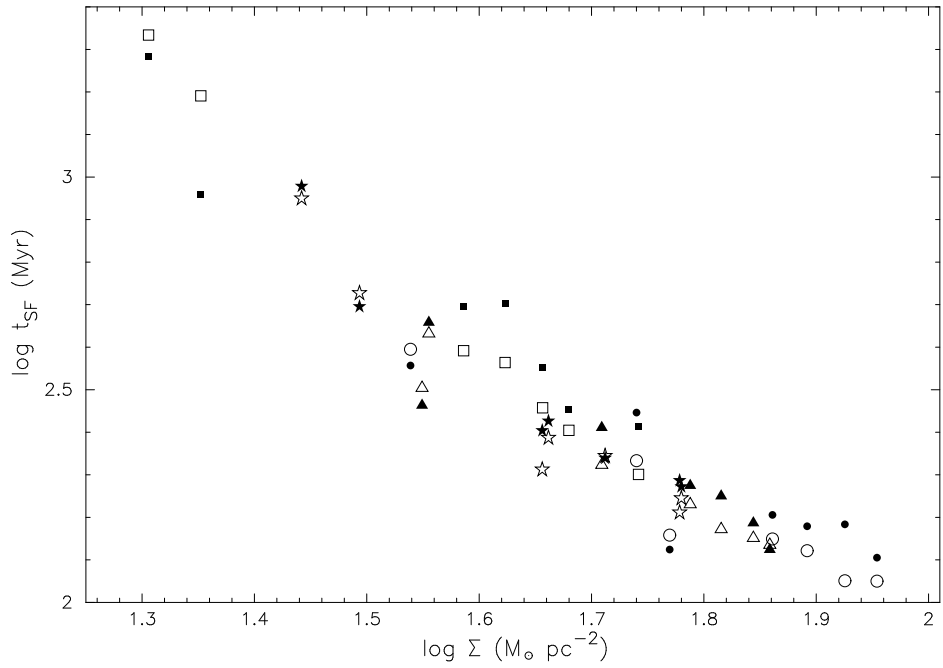}
\caption{Star formation times from model Q1D, as a function of mean
  surface density.  Different symbols correspond to different time
  bins, of width 18 Myr.  Simulation data are also binned in radii
  with widths 1 kpc.  Filled symbols show actual depletion times (eq.
  [\ref{deptime}]), and open symbols show predicted depletion times
  (eq. [\ref{deptimeapp}]), for each annular and temporal bin.  }
\label{pltdeptime}
\end{figure}

Figure \ref{pltdeptime} shows the star formation time in different radial
annuli for Model Q1D, as a function of $\Sigma$.  The actual 
times, shown by the filled symbols, are computed using equation
(\ref{deptime}), after binning the simulation data in radii of 1 kpc
widths and in time with $t/t_{orb}$ = 0.125 widths.  The open symbols
show the predicted times by applying equation
(\ref{deptimeapp}) on the same binned data.  The predicted times
agree well with the actual times.  We find similar
agreement with all other models.\footnote{Most other models do not run
  for as long as the D models that have high feedback rates, because
  some dense clumps continue to collapse without feedback, eventually
  causing the Courant condition to be violated.}  We also tested the
correlation between $t_{SF}$ and $t_{orb}$, and found no strong
correlation.  This lack of correlation occurs because at later times
the surface density profile no longer resembles the initial $R^{-1}$
profile.  

What is expected, on theoretical grounds, for the value of the star
formation time?  Consider the case in which gas cycles
between diffuse (gravitationally unbound) and dense (gravitationally
bound) components.  The diffuse component forms dense clouds at a rate
$M_{diff}/t_{diff}$, and the dense clouds are returned to the diffuse
component plus stars over a cloud lifetime at a rate
$M_{dense}/t_{cl}$.  Here $M_{diff}$ and $M_{dense}$ are the total
diffuse and dense gas masses in an annulus, with corresponding surface
densities when averaged over the area of $\Sigma_{diff}$ and
$\Sigma_{dense}$ (the latter is not to be confused with the surface
density of an individual dense cloud, which is much higher).
Similarly, $\Sigma$ is the surface density corresponding to the total
mass of all the gas $M_{tot}$ in an annulus.  In equilibrium, the
rates into and out of the dense component are equal, so that the star
formation rate per unit area averaged over the annulus is
\begin{equation}
\Sigma_{SFR}=\epsilon_{SF} \frac{\Sigma}{t_{diff}+t_{cl}}=
\epsilon_{SF} \frac{\Sigma_{diff}}{t_{diff}}=
\epsilon_{SF} \frac{\Sigma_{dense}}{t_{cl}}=
R_{SN} M_{SN} \Sigma_{dense}.
\label{rateeq}
\end{equation}
Including all the gas, the star formation timescale using the
definition of equation (\ref{deptimeapp}) (and dropping the prime) is then 
\begin{equation}
t_{SF}=\frac{t_{diff}+t_{cl}}{\epsilon_{SF}}=
\frac{t_{diff}}{\epsilon_{SF}}\times\frac{M_{tot}}{M_{diff}}=
\frac{t_{cl}}{\epsilon_{SF}}\times\frac{M_{tot}}{M_{dense}}.
\label{SFtime}
\end{equation}
Since $t_{cl}/\epsilon_{SF}=(R_{SN} M_{SN})^{-1}= 5\times10^{7}$ or
$5\times10^{8}\ \yr$ is held constant within any given model, the star
formation time for all gas in an annulus is inversely proportional to
the fraction of the gas above the density threshold in that annulus.
If most of the gas is diffuse (as is true in our simulations), then
$t_{diff}\gg t_{cl}$ and $t_{SF}\sim t_{diff}/\epsilon_{SF}$; the star
formation time is set by the typical time required for diffuse gas to
collect into bound clouds.

What characteristic values might be predicted for the cloud formation
timescale, $t_{diff}$?  The shortest possible timescale would be that
associated with the fastest-growing Jeans modes in a disk.  For a disk
with semi-thickness $H$ and sound speed $c_s$, the approximate
dispersion relation for in-plane modes is $\omega^2=k^2 c_s^2 - 2 \pi
G \Sigma |k|/(1+|k|H)$ \citep[][Paper I]{KOStone02}. For the
fastest-growing modes (which satisfy $d|\omega^2|/dk=0$) and for $H<
c_s^2/(\pi G \Sigma)$ (i.e. less than the thickness of an isothermal
disk bound only by its own gravity), the inverse of the growth rate is
$0.3-0.5 t_J$, where $t_J= c_s/(G \Sigma)$ is the thin-disk Jeans
length divided by $c_s$.  In reality, rotation, shear, and turbulence
must all affect the cloud growth timescale (see below), but the Jeans
time nevertheless provides a useful reference value.

Another reference value for a structure formation timescale that is
frequently used is the free-fall time,
$t_{ff}=(3\pi/32G\rho)^\frac{1}{2}$.  If the surface density and
volume density are related via $\Sigma = \rho H \sqrt{2 \pi}$ (as for
a Gaussian density distribution), then $t_{ff}=(3 \sqrt{2}
\pi^{3/2}H/32G\Sigma)^\frac{1}{2}$.  For our ``thick-disk'' Poisson
solver, $H\propto R$ is adopted, so that $t_{ff} \propto
(R/\Sigma)^{1/2}$.  Our initial profiles follow $R\propto
\Sigma^{-1}$, so that in the initial conditions $t_{ff}\propto
\Sigma^{-1} \propto t_J$.  In particular, for the $Q=1$ case,
$t_{ff}=0.3 t_J$ everywhere initially. Over time, however, the surface
density is spatially rearranged, so that the values of $t_J$ and
$t_{ff}$ are no longer strictly proportional.

\begin{figure}
\includegraphics[angle=-90,scale=0.5]{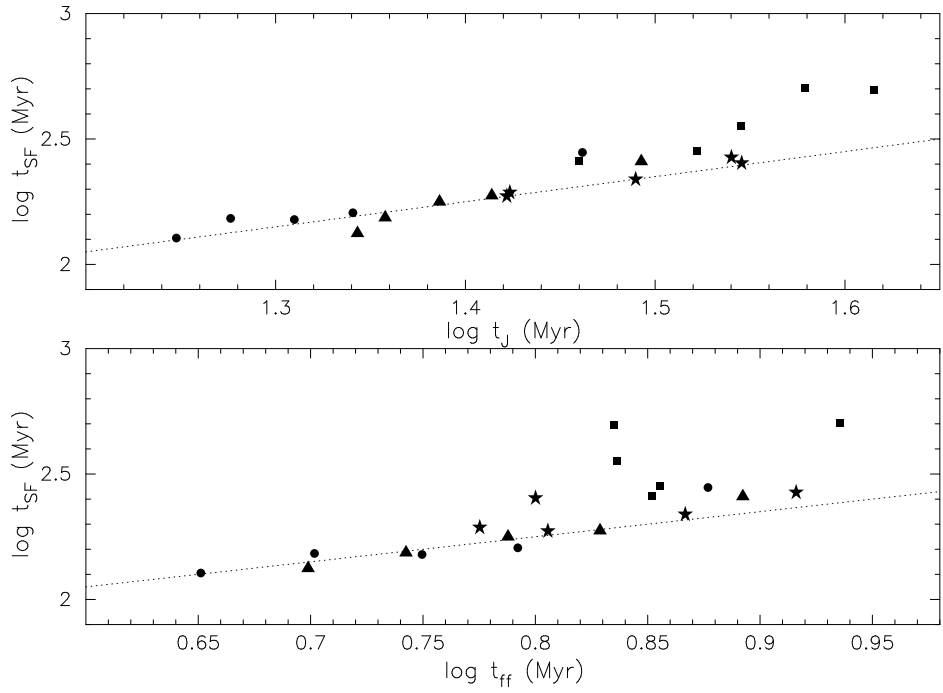}
\caption{Star formation times from model Q1D, as a function of Jeans
  time (top) and free-fall time (bottom).  Simulation data are binned
  as described in the caption to Figure \ref{pltdeptime}, with the
  innermost and outermost annuli excluded.  The dashed lines, shown
  for comparison, have slopes of 1.}
\label{t_J-t_g}
\end{figure}

Figure \ref{t_J-t_g} shows the relationships between the star
formation time and the reference values $t_J$ and $t_{ff}$.  While a
clear correlation is evident for both relations, we find that there is
less scatter in the $t_{SF}-t_J$ relation than in $t_{SF}-t_{ff}$
relation.  Further, many of the data points are consistent with a
linear relationship $t_{SF}=7t_J$, as indicated in the figure.  If we
compare to the prediction $t_{SF}=t_{diff}/\epsilon_{SF}$ and
substitute the value $\epsilon_{SF}=0.05$ used in model Q1D, this
yields $t_{diff}=0.35 t_J$, which agrees with the simple estimate
described above based on self-gravitating instabilities in thick
disks.  This result suggests that, provided the efficiencies of star
formation in GMCs are constant and the disk is dominated by diffuse
gas, the Jeans time in the diffuse gas controls the rate of star
formation.  While this result is quite intriguing, a high dynamic
range in a wider range of disk models is necessary to further
investigate this relationship.

We note that in the dispersion relation used to predict $t_{diff}\sim
t_J=c_s/(G\Sigma)$, no account was made for turbulence.  As discussed
in \S \ref{SFR_turb}, the simplest phenomenological modification of
this relation would simply be to substitute $c_s\rightarrow c_{eff}$
(see eq. [\ref{c_effeqn}]).  The results presented in \S\ref{SFR_turb}
which compare SFRs for model pairs with low and high $P_{rad}$, and
hence different $c_{eff}$, are at least semi-quantitatively in support
of this prescription for modifying $t_J$.  There, we found that an
increase of $c_{eff}$ by a factor of $\sim$ 2 is associated with a
decrease in the SFR by a factor $\sim$3.  However, the current models
are not sufficient for a definitive statement.  An important objective
for future work is to test the relation between $t_{SF}$ and the
turbulence level using a more extensive set of models; the velocity
dispersion can be varied by tuning the parameter $P_{rad}$.  A
fundamental understanding of star formation in molecular-dominated
regions of galaxies (where the thermal velocity dispersion is dwarfed
by the turbulent value) will depend on such investigations.

Modeling truly three-dimensional disks, with the vertical dimension
fully resolved, would allow for a more complete 
study of the correlations between $t_{SF}$ and the two gravitational
times, $t_J$ and $t_{ff}$.  Depending on the regime, vertical
hydrostatic equilibrium (for an isothermal medium) may be in the limit
dominated by (a) the disk's gaseous self-gravity, so that the 
effective thickness of the ISM is $\Sigma/(2\rho_0)=c_s^2/(\pi G
\Sigma)$, or (b), the disk's stellar gravity, so that the effective thickness 
is $\Sigma/(2\rho_0)=c_s \sigma_* /(2\sqrt{\pi} G \Sigma_* )\propto 
(Q_*/Q)c_s^2/(\pi G \Sigma)$.  Here,
$\sigma_*$ and $\Sigma_*$ are the stellar vertical velocity
dispersion and surface density, respectively, and $Q_*$ is the Toomre
parameter for the stellar disk.  Using these two forms, if gas
dominates the vertical gravity, then $t_{ff}\propto t_J$, whereas if
the stars dominate the vertical gravity, then $t_{ff}\propto t_J(Q_*/Q)^{1/2}$.
If galaxies evolve such that $Q_*/Q$ is constant, then $t_J \propto
t_{ff}$ in either case; it would then be empirically difficult to
establish whether  $t_J$ or $t_{ff}$ is more fundamental for
determining the star formation time. With explicit three dimensional
models, on the other hand, it will be possible to 
study the dependence of $t_{SF}$ on $t_J$
and $t_{ff}$ separately, with $Q_*/Q$ a tunable parameter.  This
represents a very interesting avenue for future research.

\section{Discussion and Summary \label{discsum}}

\subsection{Kennicutt-Schmidt Law in Simulations \label{ksdisc}}

The prescription we adopt for star formation in this paper implies a
constant relation between the mass (or mean surface density) of {\it dense}
gas and the rate (or mean surface density) of star formation,
$\Sigma_{SFR}= R_{SN} M_{SN} \Sigma_{dense}$.  Using this
prescription, we then test how the star formation rate scales with the
surface density of \textit{all} the gas.  We find that our simulations
are consistent with scalings $\Sigma_{SFR} \propto \Sigma^{1+p}$ for a
range of power law indices, but with significant scatter.  In part,
both the range of indices and the scatter in many of our models may
arise from transient effects, rather than describing the behavior in a
fully-developed star-forming disk.  Our simulations suggest that
measured star formation properties are subject to transient effects,
thus for meaningful theoretical predictions it is necessary for
systems to evolve well beyond the initial state.

For our strong-feedback model that most closely reaches an equilibrium
between cloud formation and destruction and has a large dynamic range
of surface density, we find a fairly tight relationship between
$\Sigma_{SFR}$ and $\Sigma$, with $1+p\sim 2$ (see Fig.
\ref{innercomp}).  This implies the fraction of dense gas follows
$M_{dense}/M_{tot}=\Sigma_{dense}/\Sigma\propto \Sigma$.  If we
interpret this in terms of cloud formation/destruction equilibrium
(cf. eq. \ref{rateeq}), with a constant mean cloud lifetime given by
equation (\ref{tcloud}), this implies a dense gas formation time
$\propto \Sigma^{-1}$.  As discussed in \S\ref{SFpredict}, our
quantitative results are generally consistent with a formation time
for dense gas $\propto t_J$ or $t_{ff}$, which vary (exactly or
approximately) $\propto \Sigma^{-1}$ in our models.

In other recent numerical work, star formation prescriptions
$\Sigma_{SFR}\propto \Sigma/t_{ff}$ have been adopted, where either
all of the gas or just high-density gas is included in the right-hand
side.  This would imply $\Sigma_{SFR} \propto \Sigma^{3/2}(G/H)^{1/2}$
for the dependence on surface density and disk thickness.  For a disk
in vertical hydrostatic equilibrium with vertical velocity dispersion
$\sigma_z$, the natural thickness varies as $H\propto
\sigma_z^2/(G\Sigma)$, which would imply $\Sigma_{SFR} \propto
\Sigma^2 G/\sigma_z$.  Thus, a vertically-resolved disk with a constant
vertical velocity dispersion would be expected to yield an index
$1+p=2$.  If the disk thickness is determined not by hydrostatic
equilibrium but in some other way, however, the resulting star
formation rate and the index in the K-S law would depend on the
numerical prescription (or physical process) that sets $H$.  In our
models, we have a flared disk $H\propto R$ and set $\Sigma\propto
R^{-1}$ in our initial conditions, which accounts for the index
$1+p\sim 2$ that we obtain.  If, on the other hand, the value of $H$
were constant in a given simulation (either by design for a
two-dimensional simulation, or as a consequence of limited spatial
resolution in a three-dimensional simulation), then the result would
be $1+p\sim 1.5$.  Thus, limited vertical resolution can potentially
artificially reduce the scaling index in the K-S relation, as measured
from numerical simulations.  A fully-resolved vertical dimension is
therefore required if the star formation prescription is to be based
on a volume density.  In practice, the resolution requirement can be
quite demanding if the disk is dominated by cold atomic or molecular
gas, since $c_s^2/(\pi G \Sigma)=4\pc (T/100{\rm K})(\Sigma/10$ \msun
$\, \pc^{-2})^{-1}$.  This also points to the necessity of incorporating
turbulent processes in three-dimensional models, since observed cold
gas is in fact dominated by turbulent rather than thermal pressure.
If these turbulent effects were not included, the disk thickness would
be unphysically small.

\subsection{Model Limitations and Future Prospects}

\subsubsection{Spiral Structure \label{spiral_turb}}

In spiral models, the external spiral potential is initially the
primary driver for enhancing the density, leading eventually to the
growth of clouds.  In models that evolve for a significant amount of
time, soon after feedback and the dispersal of cloud gas the global
spiral pattern is disrupted, and eventually vanishes.  With the simple
feedback prescription that we have adopted, we were unable to simulate
a spiral galaxy in which the global spiral pattern is maintained
simultaneously as cloud gas is returned to the ISM through feedback.

If the arms truly are long lasting, then either the spiral potential
is much stronger than in our models ($F>>10\%$), and/or the real
feedback events are not as disruptive of structure on kpc scales.
Very large $F$, however, does not appear consistent with observations
of the old stellar disk \citep{RixRieke93}.  One possibility is that
realistic feedback is both gentler and less correlated than the simple
prescription of our current models, and as a consequence the spiral
arm coherence would not be destroyed by large-scale shells.  Indeed,
semi-analytic models suggest that photo-ionization may evaporate much
of the mass in a typical GMC before the pressure-driven expansion of
HII regions unbinds the whole cloud
\citep[e.g.][]{KrumholzMatznerMcKee06}.  Those models do not include
supernovae, however, which are unavoidable if a GMC survives for more
than one generation of OB stars.  Still, supernovae that are less
correlated in space and time than the extreme case we have considered
would disperse cloud gas in smaller parcels.  Less correlated energy
inputs would produce shells with diameters less than the spiral arm
thickness, and could more easily leave global spiral structure intact.
By studying how the resulting spiral morphology varies with the
correlation of feedback energy, it will be possible to place limits on
how correlated star formation is in real galaxies.

\subsubsection{Multiphase ISM}

The models discussed in this paper use the simplest possible
prescription for gas thermodynamics, which is an isothermal equation
of state.  Our adopted sound speed of $c_s = 7$ \kms\ corresponds to a
temperature of $T \sim 10^4$ K, characteristic of the warm phase of
the ISM.  We adopted this approach in order to investigate, in a
controlled fashion, various separate effects that can contribute to
the regulation of star formation.

In parallel with our simplified ISM thermodynamics, our approach to
modeling feedback from star formation is also reduced to the most
basic elements. In our models, we follow the expansion of clouds
subsequent to correlated SN events.  Of course, in a real SN event,
thermal energy is injected into the ISM, and it is the expansion of a
very hot and very diffuse bubble of gas that drives the formation of a
dense shell around it.  The cooling time in the high density shell is
short, so at late stages the isothermal approximation is adequate.
The cooling time of the hot interior of each individual bubble, and of
the hot phase of the ISM that results from merging SN remnants, is
much longer.  However, the hot phase contains only a very small
fraction of the total ISM mass.  From the point of view of most of the
mass in the ISM, the primary effect of SN is to inject momentum.  By
adopting an isothermal equation of state, and treating feedback as
providing momentum inputs, this effect is captured in an approximate
way.

A significant limitation of our models is that we do not treat the
cold (T$\sim$100 K) atomic component of the ISM explicitly.  Because
the level of turbulence in the atomic component is comparable to the
thermal velocity dispersion of warm gas \citep{HeilesTroland03}, the
effective pressure in the cold medium may be comparable to the thermal
pressure in the warm medium.  The dynamics associated with ``turbulent
pressure'' may, however, be quite different from those resulting from
micro-physical thermal pressure.  A very important direction for future
work is to study directly how large-scale gravitational instabilities
and spiral structure develop in multiphase, turbulent, cloudy gas.

Another limitation of our models is that they are two-dimensional
(although the disk flares with radius).  This constrains feedback
energy to be confined within the galaxy's midplane, and does not allow
for dynamically evolving disk thickness.  In the real ISM, correlated
SN may be important in driving the SN heated gas away from the
midplane of the galaxy into the halo, through so-called chimneys and
superbubbles \citep{NormanIkeuchi89}.  To explore the effect of this
energy loss in an approximate way, in our models we consider both a
``standard'' momentum input per SN, and a momentum input reduced by a
factor of ten.  However, the cycling of gas through the galactic halo
has other consequences as well.  After this gas is cooled in the halo,
it falls back onto the disk in the form of cloudlets
\citep[e.g.][]{JoungMacLow06}.  Even though recent simulations have
shown that the fraction of mass that is vertically driven is small
\citep[e.g.][]{deAvillezBreitschwerdt04}, the in-falling clouds may
still affect the dynamics of the disk and may also act as another
source of turbulence.

In order to accurately model disks that account for the effects of SN
heating, chimneys, superbubbles, and the return of halo gas onto the
disk, a three dimensional grid, as well as explicit treatment of
heating and cooling, are necessary.  Three dimensional simulations
will also allow us to test the sensitivity of the Kennicutt-Schmidt
slope to the disk thickness (which evolves in response to star
formation), as discussed in $\S$ \ref{ksdisc}.  These directions are
important avenues for future research.

\subsection{Summary \label{summary}}

In this paper, we consider the formation of self-gravitating
structures in global models of spiral galaxies, focusing on the
effects of star formation feedback.  Our numerical simulations adopt a
simple, isothermal treatment of the gas, and follow the flow in the
disk by integrating the hydrodynamical equations on a polar grid.  We
incorporate vertical disk thickness effects within the solution of the
Poisson equation, which assumes that the disk flares as $H\propto R$.
The feedback model treats the specific star formation rate in gas
above a given surface density threshold as a constant, $R_{SN}
M_{SN}$.  Feedback is implemented by spatially-resolved radial
momentum injection subsequent to star formation events; the momentum
injection is proportional to the number of stars formed.  In order to
explore the sensitivity of the resulting model properties to the
feedback parameters, we consider a range of specific star formation
rates, star formation efficiencies $\epsilon_{SF}$, and momentum
injection per massive star $P_{rad}$.  We analyze the ISM spatial
distribution, star formation rates, and turbulent properties of our
model disks in cases with and without an externally-imposed spiral
gravitational perturbation.

Our main findings are as follows:

(1) In models where $P_{rad}$ is comparable to the level expected from
a supernova, clouds are destroyed by star formation events and the
mean turbulence level is high.  In models where $P_{rad}$ is a factor
of ten lower, to represent inefficient feedback (e.g. if SN energy is
vented vertically rather than kept in the disk), the self-gravitating
structures that form are not destroyed by feedback, and the turbulence
levels are substantially lower.  Turbulence levels are insensitive to
the star formation rate parameter $R_{SN}$ and the overall star
formation rate, however.

(2) In models with strong feedback, expanding flows lead to collisions
of shells, which then lead to gravitational collapse of overdense
regions and further star formation events. In this sense, our models
are a concrete realization of the concept of self-propagating star
formation.  We find, however, that the {\it net} effect of feedback is
to {\it lower} the rate of star formation.  That is, when we compare
models with strong feedback (large $P_{rad}$) and weak feedback (small
$P_{rad}$), the former have lower resultant star formation rates.
Similarly, when we compare models (at large and fixed $P_{rad}$) that
have high or low feedback event rates $R_{SN}$, the fraction of dense
gas is lower when the event rate is higher.  In principle, turbulence
can either enhance collapse and star formation (by inducing shell
collisions) or suppress collapse and star formation (by breaking up
overdense regions).  Our results show that although both effects
occur, the latter dominates: star formation is in net suppressed by 
feedback.

(3) For $\Sigma\sim 10-100$ \msun\ $\pc^{-2}$, the range in $\Sigma_{SFR}$
for our simulations is similar to the range observed in normal disks.
The slope of the Kennicutt-Schmidt scaling relation $\Sigma_{SFR}
\propto \Sigma^{1+p}$ is steeper ($1+p\sim 2$) in our simulations than
the slopes found from current observations at high (average) surface density.
The discrepancy may be due to our assumption that the disk thickness
varies with radius as $H\propto R$.  Indeed, our numerical results are
consistent with the theoretical prediction that $t_{SF}\propto t_J$ or
$t_{ff}$ when the gravitational times $t_J$ and $t_{ff}$ are
calculated based on our model prescription.  We point out that
shallower scalings of $\Sigma_{SFR}$ with $\Sigma$ would be expected
if the vertical velocity dispersion increases with $\Sigma$.  This
would increase the disk thickness at small radii (where $\Sigma$ is
large) relative to what we have assumed, and consequently increase the
gravitational times and reduce $\Sigma_{SFR}$.  

(4) Motivated by our own results, we remark that in general, the
thickness of the gaseous disk in a galaxy (either observed or
simulated) is important for setting the index in the Kennicutt-Schmidt
relationship.  Numerical simulations must resolve the natural disk
scale height (set by pressure and turbulence) if the adopted
prescription for star formation depends on the volume density $\rho$
of gas.  A simulation that is vertically unresolved ($H\rightarrow
const.$) while adopting
$\rho_{SF} \propto \rho/t_{ff}(\rho)$, and hence $\Sigma_{SF}
\propto \Sigma/t_{ff}(\rho)$, will automatically yield an index
$1+p=1.5$ in the K-S law since $t_{ff}^{-1}\propto
(\Sigma/H)^{0.5}$.  Fundamental understanding of K-S laws requires a
self-consistent determination of the dependence of $H$ on $\Sigma$.

(5) For turbulence driven by expanding shells in overdense regions, we
find that the power spectra decrease with decreasing size consistent
with the scalings for shock-dominated flows (``Burgers turbulence'').
While typical mass-weighted velocity dispersions on kpc scales in our
high-$P_{rad}$ models are 31 \kms, these decrease to 10 \kms\ on 100
pc scales.  Radial and azimuthal components of the velocity dispersion
in a given scale are comparable.

(6) For all of our models, the maximum masses of dense clouds that form
are several million \msun, consistent with observations of the upper
cutoff in GMC/GMA mass distributions in local group galaxies.  In
models with strong turbulence, such that self-gravitating
condensations can form in colliding flows, a wider range of cloud
masses results, with a lower peak in the distribution (but similar
upper cutoff).  Higher resolution simulations will allow for a more
detailed analysis of the mass distributions.

(7) Within the context of the feedback prescription and parameters for
our current set of models, we find that global spiral patterns are not
maintained.  For low $P_{rad}$, insufficient momentum is injected to
overdense structures so that arm clouds continue to collapse,
eventually depleting the surrounding spiral arm gas.  For high
$P_{rad}$, large-scale expanding shells form and the global spiral
structure is destroyed as cloud gas is dispersed.  We conclude that
highly-correlated star formation, which is the limit that we adopt in
the present models, is incompatible with long-lived spiral structure.
It will be interesting to determine, by comparing spiral morphology
with results from models adopting differing feedback prescriptions,
what constraints are placed on the spatial and temporal correlation of
star formation feedback in real galaxies.

\acknowledgements 

This work was supported by grants 1278889 (NASA/{\it Spitzer}), and
AST 0507315 (NSF).  Computations were performed on clusters supported
by the Center for Theory and Computation in the Astronomy Department
at the University of Maryland.  For much of the data analysis and
visualization, we have made use of {\tt NEMO} software
\citep{Teuben95}.  We are also grateful to the anonymous referee for
helpful comments.

\appendix
\section{Appendix}
In the Appendix of Paper I, we described two methods to solve
Poisson's equation numerically on a polar grid; both methods employ
Fast Fourier Transforms (FFTs).  One method sums the potential from
concentric rings, as described by \citet{Miller76}.  The other method
employs a coordinate transformation from polar coordinates to a
Cartesian-like coordinate system.  The former method is exact, but
computationally expensive, and the latter is an approximation, but
computationally efficient.

Here, we describe another FFT based method that is exact, and more
efficient than the \citet{Miller76} method.\footnote{As in Paper I, we
  again make use of the freely-available FFTW software
  \citep{FFTW05}.}  
The basic scheme is described in \citet{Kalnajs71}
and \citet{BT87}; we describe a modification of Kalnajs's method that
includes the effect of nonzero disk thickness $H$, which also acts as
softening.

The potential $\Phi$ at each position
$(R,\phi)$ on the disk, at $z$=0, is

\begin{equation}
\Phi(R,\phi,z=0) = -G \int dR' \int d\phi' \int dz' \frac{R' f(z',R',\phi')\Sigma(R',\phi')}{[R^{'2}+R^2-2RR'\cos(\phi-\phi')+z^{'2}]^\frac{1}{2}}.
\label{phi}
\end{equation}
Here, $G$ is the usual gravitational constant, $\Sigma$ is the total
surface density,  and the 
function $f=\rho(z',R',\phi')/\Sigma(R',\phi')$
describing the vertical profile of the volume 
density must be normalized,
$\int_{-\infty}^{\infty}dz' f(z',R',\phi') = 1$.  
Substituting $u'\equiv \ln R'$, and $\zeta' =
z'/\sqrt{2}R'$ in equation (\ref{phi}), the potential reduces to
\begin{equation}
\Phi(R,\phi,z=0) = -Ge^u \int du' \int d\phi' \int d\zeta'
\frac{ e^{u'-u} e^{u'} f(\zeta',u',\phi')\Sigma(u',\phi')}
{[e^{u-u'}(\cosh(u-u')-\cos(\phi-\phi'))+\zeta^{'2}]^\frac{1}{2}}.
\label{phi_sub}
\end{equation}
If $R' \rho(z',R',\phi')/\Sigma(R',\phi')= 
e^{u'}f(\zeta',u',\phi')\equiv g(\zeta')$ 
is a function of
$\zeta'$ only (see below), we
can define 
\begin{equation}
I(u'-u,\phi'-\phi) \equiv e^{u'-u}\int d\zeta' 
\frac{ g(\zeta')}{[e^{u-u'}(\cosh(u-u')-\cos(\phi-\phi'))+\zeta^{'2}]^\frac{1}{2}}.
\label{Idef}
\end{equation}
Using the definition of $I$ in equation (\ref{phi_sub}),
we obtain $\Phi$ as a two dimensional convolution:
\begin{equation}
\Phi(R,\phi,z=0) = -Ge^u \int du' \int
d\phi'\Sigma(u',\phi')I(u'-u,\phi'-\phi).
\label{phi_noz}
\end{equation}
Applying the Fourier convolution theorem to equation 
(\ref{phi_noz}), the gravitational potential can
be computed by taking the Fourier transform of $\Sigma$ to obtain
$\hat\Sigma$, and then taking the inverse Fourier transform of the
product of $\hat\Sigma$ and $\hat I$, where $\hat I$ is the Fourier
transform of $I$.  In hydrodynamic simulations, $\hat I$ can be
computed once at the beginning of the simulation run, so that only two
FFTs need to be performed at each timestep, FFT($\Sigma$) and
FFT$^{-1}$($\hat\Sigma \hat I$).

The function $I$, and therefore its convolution $\hat I$, depends on
the normalized vertical distribution function $g(\zeta)$.
For the specific case of a Gaussian vertical density distribution
(which holds if the vertical gravity is dominated by that of the
stellar disk), 
\begin{equation}
f(z^{},R^{}) = \frac{e^{-z^{2}/2H^2(R^{})} }{\sqrt{2\pi H^2(R^{})}}.
\label{fgaus}
\end{equation}
For a disk that flares as $H(R^{}) \propto R^{}$, we define
${\cal H} = H(R^{})/R^{}$, so that
\begin{equation}
e^{u^{}}f(\zeta^{},R^{}) = \frac{e^{-(\zeta^{}/{\cal
      H})^2}}{\sqrt{2\pi}{\cal H}}\equiv g(\zeta^{}).
\label{fthick}
\end{equation}
Similarly, if the vertical density follows a ${\rm sech}^2$ distribution
(true if the gaseous self-gravity dominates), then
$g(\zeta^{})=(2{\cal H})^{-1}{\rm sech}^2(\zeta^{}\sqrt{2}/{\cal
  H})$.

For our simulations, we adopt the Gaussian profile; this yields the following
explicit expression for $I$:
\begin{equation}
I(u'-u,\phi'-\phi) \equiv \frac{e^{u'-u}}{\sqrt{2\pi}{\cal
    H}} \int_{-\infty}^{\infty} d\zeta' 
\frac{ e^{-(\zeta'/{\cal H})^2}}{[e^{u-u'}(\cosh(u-u')-\cos(\phi-\phi'))+\zeta^{'2}]^\frac{1}{2}}.
\label{Iexp}
\end{equation}

Finally, we comment on the assumption $H(R)\propto R$ which enables
the three-dimensional gravitational integral to be written as a
two-dimensional convolution.  If the stellar disk dominates gravity,
then for an isothermal disk the vertical density distribution is
Gaussian with $H/R = c_s Q_* (c_{*,z}/c_{*,R})/2v_c$, so values of
$c_s/v_c$, $Q_*$, and $c_{*,z}/c_{*,R}$ that are independent of radius
imply constant $H/R$.  Similarly, if gas is the dominant component for
vertical gravity, $H = c_s^2/\pi G \Sigma$, so that $H/R = c_s
Q/\sqrt{2}v_c$.  If both the Toomre $Q$ parameter and $c_s/v_c$ are
independent of $R$, then $H/R$ = constant.  For self-gravitating
gaseous disks, if $Q=1$ and $v_c/c_s$=30, then $H/R = 0.02$.
Including stellar gravity typically reduces $H$ by a factor of $\sim2$
\citep[e.g][]{KOStone02}.

For the simulations described in this paper, we use $H/R\equiv {\cal
  H}=0.01$ in equation (\ref{Iexp}).  We have tested other values of
$\cal H$, and find that our results are not sensitive to the exact
value.  However, large changes significantly affect the rate of
growth of self-gravitating perturbations.

\bibliography{ref}

\begin{deluxetable}{cccccc} 
  \tablewidth{0pt} \tablecaption{Parameters for
    Standard\tablenotemark{a} Models} \tablehead{ \colhead{Model} &
    \colhead{$Q_0$} & \colhead{$m$} &
    \colhead{$R_{SN}$\tablenotemark{b} } & \colhead{$\epsilon_{SF}$}
    & \colhead{$P_{rad}$ ($10^5$ \Mkms)} \\
    \colhead{(1)} & \colhead{(2)} & \colhead{(3)} & \colhead{(4)} &
    \colhead{(5)} & \colhead{(6)} } \startdata
  Q1A & 1 & 0 & 1 & 0.025 & 0.34  \\
  Q1B & 1 & 0 & 1 & 0.05 & 0.34  \\
  Q1D & 1 & 0 & 10 & 0.05 & 3.4  \\
  Q1SA & 1 & 4 & 1 & 0.025 & 0.34  \\
  Q1SB & 1 & 4 & 1 & 0.05 & 0.34  \\
  Q1SC & 1 & 4 & 1 & 0.05 & 3.4  \\
  Q1SD & 1 & 4 & 10 & 0.05 & 3.4 \\
  Q1SE & 1 & 4 & 10 & 0.05 & 0.34  \\
  Q2SA & 2 & 4 & 1 & 0.025 & 0.34  \\
  Q2SB & 2 & 4 & 1 & 0.05 & 0.34  \\
  Q2SC & 2 & 4 & 1 & 0.05 & 3.4  \\
  Q2SD & 2 & 4 & 10 & 0.05 & 3.4  \\
  \enddata {\singlespace \tablenotetext{a}{\footnotesize
      1024$\times$1024 zones; R $\in$ 4-11 kpc; $\phi \in$
      0-$\frac{\pi}{2}$ radians} \tablenotetext{b}{\footnotesize
      Units of $2\times 10^{-11} {\rm M}_\odot^{-1}\, {\rm yr}^{-1}$, i.e.
      number of SN per 50 years per $10^9$M$_\odot$} }
\label{standardmods}
\end{deluxetable}

\begin{figure}
\includegraphics[angle=0,scale=0.75]{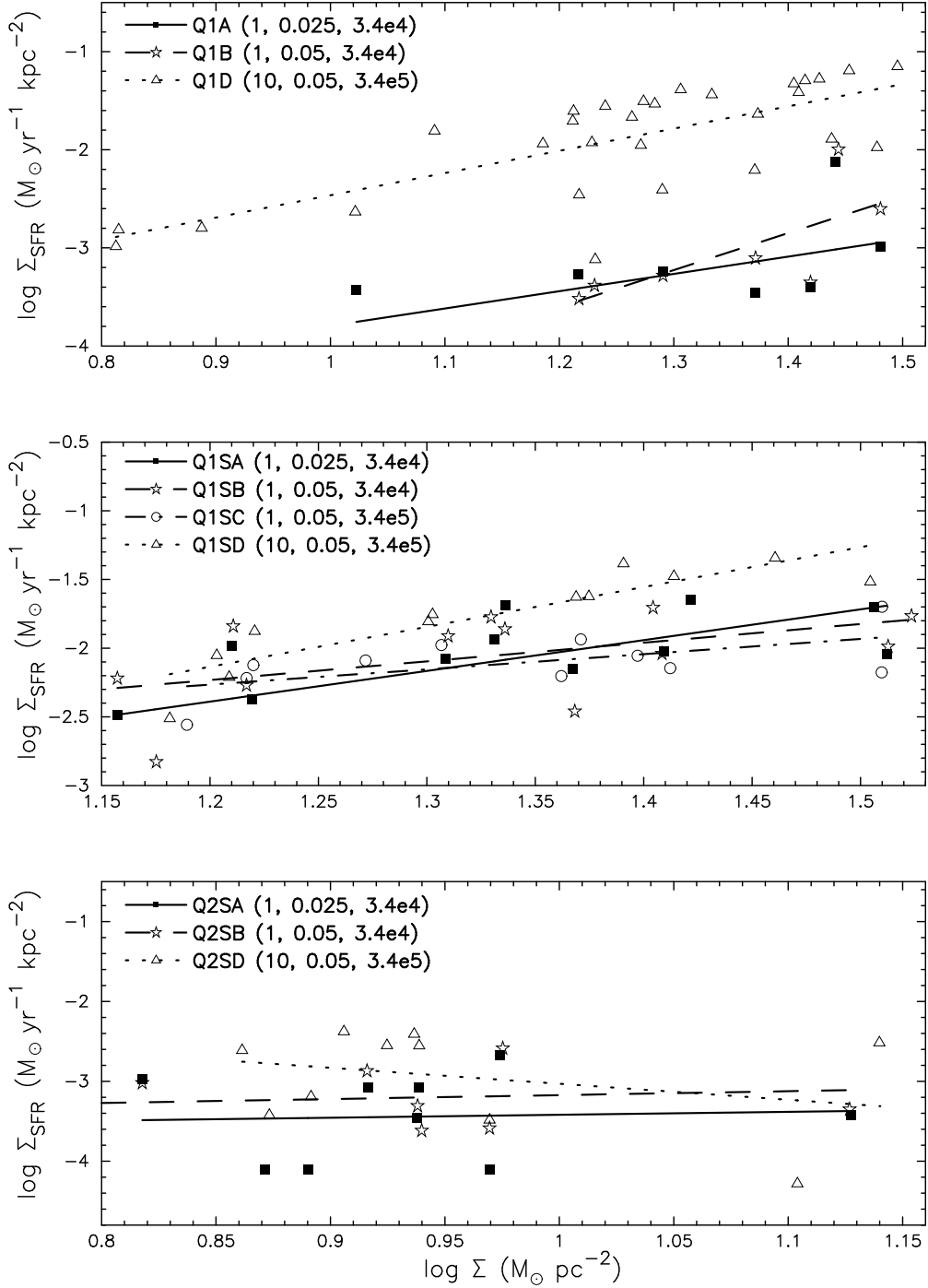}
\caption{Schmidt law for models in Table \ref{standardmods}.  Each
  point is obtained by binning the simulation data in radii, with 1
  kpc widths, and in time, with 18 Myr widths.  Only annuli and times
  with at least 1 feedback event are included.  The lines are the best
  fit to the points; with their slopes ranging from 1.8 - 3.8 (top),
  1.1 - 3.0 (middle), and -2.0 - 1.9 (bottom).}
\label{QKS}
\end{figure}

\end{document}